\documentclass{aa}  

\usepackage{graphicx}
\usepackage{multirow}
\usepackage{txfonts}
\newcommand{\Msun}{\ensuremath{\,{M}_\odot}}                            % Solar mass symbol
\newcommand{\Rsun}{\ensuremath{\,{R}_\odot}}                            % Solar radius symbol
\newcommand{\Teff}{\ensuremath{T_{\rm eff}}}                            % Effective temperature symbol
\newcommand{\logg}{\ensuremath{\log g}}                                 % log(g) symbol
\newcommand{\kms}{\,km\,s$^{-1}$}                                       % km/s symbol
\newcommand{\ms}{\,m\,s$^{-1}$}                                         % m/s^2 symbol
\begin{document}

   \title{The highly inflated giant planet WASP-174b}

   \author{
L. Mancini\inst{1,2,3}\fnmsep\thanks{lmancini@roma2.infn.it}
\and
P. Sarkis\inst{2}
\and
Th. Henning\inst{2}
\and
G.~\'{A}. Bakos\inst{4}
\and
D. Bayliss\inst{5}
\and
J. Bento\inst{6}
\and
W. Bhatti\inst{4}
\and
R. Brahm\inst{7,8,9}
\and
Z. Csubry\inst{4}
\and
N. Espinoza\inst{2}
\and
J. Hartman\inst{4}
\and
A. Jord\'{a}n\inst{10,9,8}
\and
K. Penev\inst{11}
\and
M. Rabus\inst{8,2,12,13}
\and
V. Suc\inst{8}
\and
M. de Val-Borro\inst{14}
\and
G. Zhou\inst{15}
\and
G. Chen\inst{16,17}
\and
M. Damasso\inst{3}
\and
J. Southworth\inst{18}
\and
T.~G. Tan\inst{19}
}
\institute{
%1
Department of Physics, University of Rome ``Tor Vergata'', Via della Ricerca Scientifica 1, I-00133, Rome, Italy
%\email{lmancini@roma2.infn.it}
\and
%2
Max Planck Institute for Astronomy, K\"{o}nigstuhl 17, D-69117, Heidelberg, Germany
\and
%3
INAF -- Turin Astrophysical Observatory, via Osservatorio 20, I-10025, Pino Torinese, Italy
\and
%4
Department of Astrophysical Sciences, Princeton University, NJ 08544, USA
\and
%5
Department of Physics, University of Warwick, Coventry CV4 7AL, UK
\and
%6
Research School of Astronomy and Astrophysics, Australian National University, Canberra, ACT 2611, Australia
\and
%7
Center of Astro-Engineering UC, Pontificia Univ. Cat\'{o}lica de Chile, Av. Vicu\~{n}a Mackenna 4860, 7820436 Macul, Santiago, Chile
\and
%8
Instituto de Astrof\'{i}sica, Pontificia Universidad Cat\'{o}lica de Chile, Av. Vicu\~{n}a Mackenna 4860, 7820436 Macul, Santiago, Chile
\and
%9
Millenium Institute of Astrophysics, Av. Vicu\~{n}a Mackenna 4860, 7820436 Macul, Santiago, Chile
\and
%10
Facultad de Ingenier\'ia y Ciencias, Universidad Adolfo Ib\'a\~nez, Av.\ Diagonal las Torres 2640, Pe\~nalol\'en, Santiago, Chile
\and
%11
Department of Physics, University of Texas at Dallas, Richardson, TX 75080, USA
\and
%12
Las Cumbres Observatory Global Telescope, 6740 Cortona Dr., Suite 102, Goleta, CA 93111, USA
\and
%13
Department of Physics, University of California, Santa Barbara, CA 93106-9530, USA
\and
%14
Astrochemistry Laboratory, Goddard Space Flight Center, NASA, 8800 Greenbelt Rd., Greenbelt, MD 20771, USA
\and
%15
Harvard-Smithsonian Center for Astrophysics, 60 Garden St., Cambridge, MA 02138, USA
\and
%16
Key Laboratory of Planetary Sciences, Purple Mountain Observatory, Chinese Academy of Sciences, Nanjing 210008, China
\and
%17
Instituto de Astrof\'{i}sica de Canarias, V\'{i}a L\'{a}ctea s/n, E-38205 La Laguna, Tenerife, Spain
\and
%18
Astrophysics Group, Keele University, Keele ST5 5BG, UK
\and
%19
Perth Exoplanet Survey Telescope, Perth, Australia
}

   \date{Received ; accepted }

% \abstract{}{}{}{}{} 
% 5 {} token are mandatory 
  \abstract
   {The transiting exoplanetary system WASP-174 was reported to be composed by a main-sequence F star ($V=11.8$\,mag) and a giant planet, WASP-174b (orbital period $P_{\rm orb}=4.23$\,days).  However only an upper limit was placed on the planet mass ($<1.3\, M_{\rm Jup}$), and a highly uncertain planetary radius ($0.7-1.7\, R_{\rm Jup}$) was determined.}
   {We aim to better characterise both the star and the planet and precisely measure their orbital and physical parameters. 
%and investigate the limb-darkening effect of the parent star and the variation of the radius of its planet both as a function of the %wavelength.
}
   {In order to constrain the mass of the planet, we obtained new measurements of the radial velocity of the star and joined them with those from the discovery paper. Photometric data from the HATSouth\thanks{The HATSouth network is operated by a collaboration consisting of Princeton University (PU), the Max Planck Institute for Astronomy (MPIA), the Australian National University (ANU), and the Pontificia Universidad Cat\'{o}lica de Chile (PUC). The station at Las Campanas Observatory (LCO) of the Carnegie Institute is operated by PU in conjunction with PUC, the station at the High Energy Spectroscopic Survey (H.E.S.S.) site is operated in conjunction with MPIA, and the station at Siding Spring Observatory (SSO) is operated jointly with ANU. Based on observations made with the MPG\,2.2\,m Telescope at the ESO Observatory in La Silla. Based in part on observations made with TESS, the facilities of the Las Cumbres Observatory Global Telescope, the Perth Exoplanet Survey Telescope, and the 3.9\,m Anglo-Australian Telescope at SSO.} survey and new multi-band, high-quality (precision reached up to 0.37~mmag) photometric follow-up observations of transit events were acquired and analysed for getting accurate photometric parameters. We fit the model to all the observations, including data from the TESS space telescope, in two different modes: incorporating the stellar isochrones into the fit, and using an empirical method to get the stellar parameters. The two modes resulted to be consistent with each other to within $2~\sigma$.}
   {We confirm the grazing nature of the WASP-174b transits with a confidence level greater than $5~\sigma$, which is also corroborated by simultaneously observing the transit through four optical bands and noting how the transit depth changes due to the limb-darkening effect. We estimate that $\approx76\%$ of the disk of the planet actually eclipses the parent star at mid-transit of its transit events.
We find that WASP-174b is a highly-inflated hot giant planet with a mass of $M_{\rm p}=0.330\pm0.091\,M_{\rm Jup}$ and a radius of $R_{\rm p}=1.435\pm0.050 \,R_{\rm Jup}$, and is therefore a good target for transmission-spectroscopy observations. With a density of $\rho_{\rm p}=0.135 \pm 0.042$\,g\,cm$^{-3}$, it is amongst the lowest-density planets ever discovered with precisely measured mass and radius.}
   {}

   \keywords{planetary systems -- stars: fundamental parameters -- stars: individual: WASP-174 -- techniques: photometric -- techniques: radial velocities - method: data analysis
               }

   \maketitle
%
%-------------------------------------------------------------------

\section{Introduction}

After almost 20 years of initially pioneering and then systematic operations, ground-based surveys, based on single or on an array of small automated telescopes or telephoto lenses, have made a huge contribution to the discovery of exoplanets. They open the way to current and next generation space-telescope surveys for transiting exoplanets. Nevertheless, the work of robotic telescopes is far from finished as they are exceptionally well-suited to detecting hot giant planets. Given that hot giant planets are known to be relatively rare from both RV surveys \citep{mayor:2011,wright:2012} and transit surveys \citep{fressin:2013, fulton:2018}, in order to discover such planets it is necessary to scan the sky with wide-field instruments that are sensitive to various ranges of stellar magnitudes. For this purpose, nearly 10-year long observational campaigns have been carried out by ground based observatories, equipped with small automated telescopes \citep[e.g.][]{bakos:2004,pollacco:2006,pepper:2007}. More recently, this approach of multiple, wide-field, small aperture telescopes has been adopted from a space environment with the successful TESS\footnote{Transiting Exoplanet Survey Satellite.} mission \citep{ricker:2015}.  

The formation, evolution and possible migration of hot giant exoplanets are complex phenomena, due to many interacting astrophysical processes. Observational constraints are thus necessary to discriminate the best theoretical models among the competing ones that have been proposed. A large sample of giant exoplanets, whose physical properties are well known, is needed for obtaining good statistics, and very accurate measurements of their mass and radius is crucial for this purpose. However, the complete characterization of an exoplanetary system requires, in many cases, additional observational data and modeling.

Recently, the WASP team announced the discovery of a transiting exoplanet orbiting the star WASP-174 \citep{temple:2018}, whose main parameters are summarised in Table~\ref{tab:stellarparameters}. This planetary system was reported as composed by a F6\,V star and a near-grazing transiting planet, which rotates on a moderately misaligned and (assumed) circular orbit in $4.23$\,days, but with only an upper limit on the mass ($M_{\rm p}<1.3\, M_{\rm Jup}$) due to the relatively low precision of the radial velocity measurements over most of the orbital phase ($>50$\,m\,s$^{-1}$). Also the radius of the planet was difficult to constrain precisely ($0.7 < R_{\rm p}/R_{\rm Jup} < 1.7$), because the transit is either grazing or near-grazing and does not show the second and third contact points. The planetary nature of the planet was in fact confirmed by taking a continuous series of spectra during a transit event with the HARPS high-resolution spectrograph and performing the Doppler tomographic method \citep{temple:2018}.

\begin{table*}
\caption{Astrometric, photometric and spectroscopic parameters for WASP-174.}
\label{tab:stellarparameters}      
\centering  
\begin{tabular}{lccc}
\hline
Quantity & {This work} & Source \\
\hline  \\[-6pt]%%
\multicolumn{1}{l}{\textbf{Cross-identifications}} \\ [2pt]
~~~~2MASS-ID\dotfill & {13031055-4123053}    & 2MASS     \\
~~~~GSC-ID\dotfill   & {7781-00928}          & GSC       \\
~~~~GAIA-ID\dotfill  & {6139403446074853376} & GAIA DR1  \\
~~~~GAIA-ID\dotfill  & {6139403450370668800} & GAIA DR2  \\ 
~~~~TIC\dotfill  & {102192004} & TESS  \\ [4pt]
\multicolumn{1}{l}{\textbf{Astrometric properties}} \\ [2pt]
~~~~R.A. (J2000)\dotfill & {$13^{\rm h}\,03^{\rm m}\,10.563^{\rm s}$} & GAIA DR2 \\
~~~~Dec. (J2000)\dotfill & {$-41^{\circ}\,23{\arcmin}\,05.412{\arcsec}$} & GAIA DR2  \\
~~~~$\mu_{\rm R.A.}$ (mas yr$^{-1}$)\dotfill & {\ensuremath{0.043\pm0.071}} & GAIA DR2  \\
~~~~$\mu_{\rm Dec.}$ (mas yr$^{-1}$)\dotfill & {\ensuremath{-5.784\pm0.112}} & GAIA DR2  \\ 
~~~~Parallax  (mas) \dotfill & {$2.411 \pm 0.060$} & GAIA DR2  \\ [4pt]
\multicolumn{1}{l}{\textbf{Photometric properties}} \\ [2pt]
~~~~$G$ (mag)\dotfill & {\ensuremath{11.6826 \pm 0.0003}} & GAIA DR2  \\
~~~~$BP$(mag)\dotfill & {\ensuremath{11.9495\pm 0.0009}} & GAIA DR2  \\
~~~~$RP$(mag)\dotfill & {\ensuremath{11.2753 \pm 0.0010}} & GAIA DR2  \\ [2pt]
~~~~$B$ (mag)\dotfill & {\ensuremath{12.30\pm0.10}} & APASS  \\
~~~~$V$ (mag)\dotfill & {\ensuremath{11.790\pm0.060}} & APASS \\
~~~~$g$ (mag)\dotfill & {\ensuremath{12.00\pm0.10}} & APASS  \\
~~~~$r$ (mag)\dotfill & {\ensuremath{11.677\pm0.090}} & APASS  \\
~~~~$i$ (mag)\dotfill & {\ensuremath{11.56\pm0.12}} & APASS  \\ [2pt]
~~~~$J$ (mag)\dotfill & {\ensuremath{10.845\pm0.027}} & 2MASS  \\
~~~~$H$ (mag)\dotfill & {\ensuremath{10.599\pm0.024}} & 2MASS  \\
~~~~$K$ (mag)\dotfill & {\ensuremath{10.579\pm0.021}} & 2MASS  \\ 
~~~~$W1$ (mag)\dotfill & {\ensuremath{10.530\pm0.022}} & WISE  \\ 
~~~~$W2$ (mag)\dotfill & {\ensuremath{10.561\pm0.021}} & WISE  \\ 
~~~~$W3$ (mag)\dotfill & {\ensuremath{10.510\pm0.064}} & WISE  \\ 
~~~~$W4$ (mag)\dotfill & {\ensuremath{8.964}} & WISE  \\ [4pt]
\multicolumn{1}{l}{\textbf{Spectroscopic properties}} \\ [2pt]
~~~~\Teff$_{\star}$ (K)\dotfill & {\ensuremath{6400 \pm 100}} & \citet{temple:2018} \\
~~~~[Fe/H]\dotfill & {\ensuremath{0.090\pm0.090}} & \citet{temple:2018} \\
~~~~$v \sin{i}$ (\kms)\dotfill & {\ensuremath{16.50\pm0.50}} & \citet{temple:2018}  \\
\hline
\end{tabular}
\end{table*}

Similar to the case of WASP-67b \citep{hellier:2012}, WASP-174b seems to satisfy the criterion for a grazing transiting planet, that is $b+(R_{\rm p}/R_{\star})>1$, where $b$ is the impact parameter. In a such grazing orbit, the second and third contact points are missing and the transit light curve has a V shape, similar to that of an eclipsing binary. Consequently the photometric parameters, like $R_{\rm p}/R_{\star}$ and $b$, are difficult to constrain and we need very high-quality light curves to reduce the uncertainties to levels similar to those of the other well-studied, non-grazing, transiting exoplanets ($\lesssim 20\%$).

WASP-174b was detected independently as a candidate transiting planet from HATSouth survey\footnote{\url{http://hatsouth.org/}}, and thus we have observations dating back to 2011 for discovery photometry and 2014 for follow-up spectroscopy and photometry.  Here, we use these observations, in conjunction with other data, including TESS data, to perform a comprehensive study of the WASP-174 system, which allows us to much better characterize the physical parameters of this planetary system. 

% Observations
%-------------------------------------------------------------------
\section{Observations}
\label{sec:observations}
% 
%-------------------------------------------------------------------
\subsection{The HAT South survey}
\label{sec:hatsouth}
%-------------------------------------------------------------------
HATSouth \citep{bakos:2013:hatsouth} is a three-station network of ground-based, automated wide-field telescopes. The stations are placed in the southern hemisphere, on three different continents (South America, Africa and Australia). Each station hosts two units composed of a single mount, on which four 18\,cm Takahashi astrographs are placed; 24 telescopes in total. Each of the six units can monitor a region on the celestial sphere of $8^{\circ} \times 8^{\circ}$. The large separation in longitude of the three stations allows us to monitor the same stellar field for 24\,hr per day without interruptions, except for bad weather.
The images are recorded by Apogee U16M Alta CCD cameras, automatically calibrated and then stored in the HATSouth archive at Princeton University. Aperture photometry is performed for extracting light curves, which are then analysed with the aim to find periodic signals due to transiting exoplanets. Finally, the most promising candidates undergo spectral reconnaissance, radial-velocity (RV) measurements and precise photometric follow-up observations with larger telescopes to confirm they are real exoplanets. The entire process is exhaustively described, step by step, in \citet{bakos:2013:hatsouth} and summarised in most of the previous works of the HATSouth team\footnote{The complete list of the HATSouth discovery papers is available on \url{http://hatsouth.org/planets/}}.

%
%-------------------------------------------------------------------
\subsection{Photometric detection}
\label{sec:detection}
\begin{figure}
\centering
\includegraphics[width=\hsize]{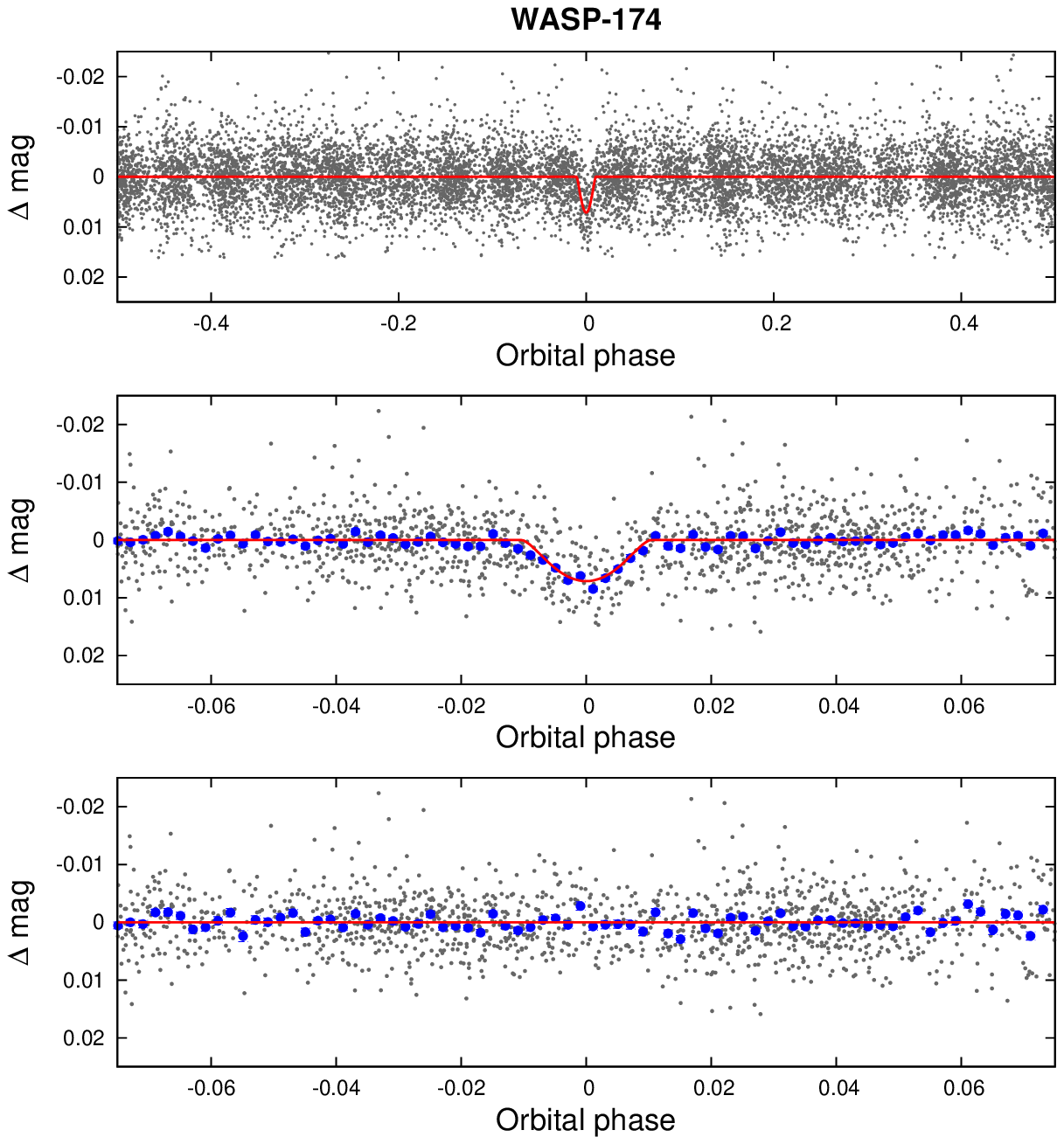}
\caption{Phase-folded unbinned HATSouth light curves for WASP-174. The top panel shows the full light curve, while the middle panel shows the light curve zoomed-in on the transit. The solid line shows the model fit to the light curve. The residuals of the fit are shown in the bottom panel. The dark filled circles in the bottom panels show the light curves binned in phase with a bin size of 0.002.}
\label{fig:hats_lc}
\end{figure}

%-------------------------------------------------------------------
WASP-174 (aka 2MASS 13031055-4123053, aka HATS700-010) was observed, through a Sloan-$r$ filter, by the HATSouth telescopes between April 2011 and July 2012 (see Table~\ref{tab:photobs_1} for details).
The subsequent analysis of its light curve, which is plotted in Figure~\ref{fig:hats_lc}, resulted in a typical transiting-planet signal with a period of $\approx 4.23$\,days. WASP-174 was thus flagged as planetary-system candidate and underwent follow-up observations for confirming its planetary nature. 
%--------------------
%--------------------
\begin{table*}
\caption{Summary of photometric observations from the HATSouth survey.}
\label{tab:photobs_1}      
\centering                          
\begin{tabular}{lccccc}        
\hline\hline  \\ [-6pt]                
Instrument/Field & Date(s) & \# Images & Cadence & Filter & Precision \\
 &  & & (s) &  & (mmag) \\
\hline \\ [-6pt]                        
~~~~HS-2.2/G700 & 2011 Apr 26 -- 2012 Jul 30 & 4409 &  321.3& $r$ & 5.6 \\
~~~~HS-4.2/G700 & 2011 Jul 8 -- 2012 Jul 29 & 3815 & 328.5 & $r$ & 4.9 \\
~~~~HS-6.2/G700 & 2012 Jan 5 -- 2012 Jul 31 & 1464 & 303.2 & $r$ & 5.2 \\
\hline                                   
\end{tabular}
\end{table*}  
%--------------------
%
%-------------------------------------------------------------------
\subsection{Spectroscopic observations}
\label{sec:obsspec}
\begin{figure}
\centering
\includegraphics[width=\hsize]{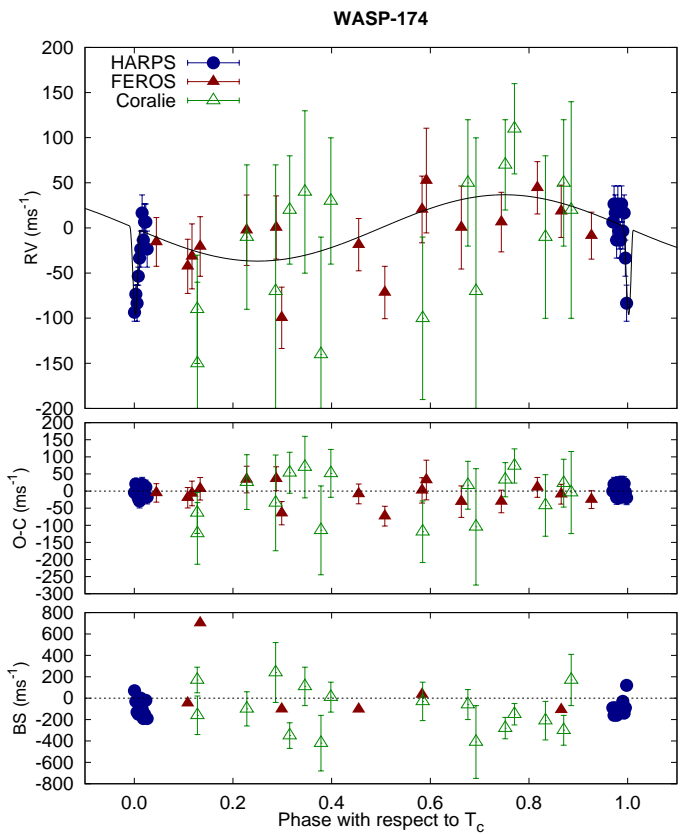}
\caption{Phased high-precision RV measurements for WASP-174. The instruments used are labelled in the plots. The top panel shows the phased measurements together with our best-fit model. Zero-phase corresponds to the time of mid-transit. The center-of-mass velocity has been subtracted. The solid line shows the best-fitting model (see discussion in Sect.~\ref{sec:analysis}). The second panel shows the velocity O-C residuals from the best fit. 
The third panel shows the bisector spans (BS). Note the different vertical scales of the panels.}
\label{fig:RV}
\end{figure}

\begin{figure}
\centering
\includegraphics[width=\hsize]{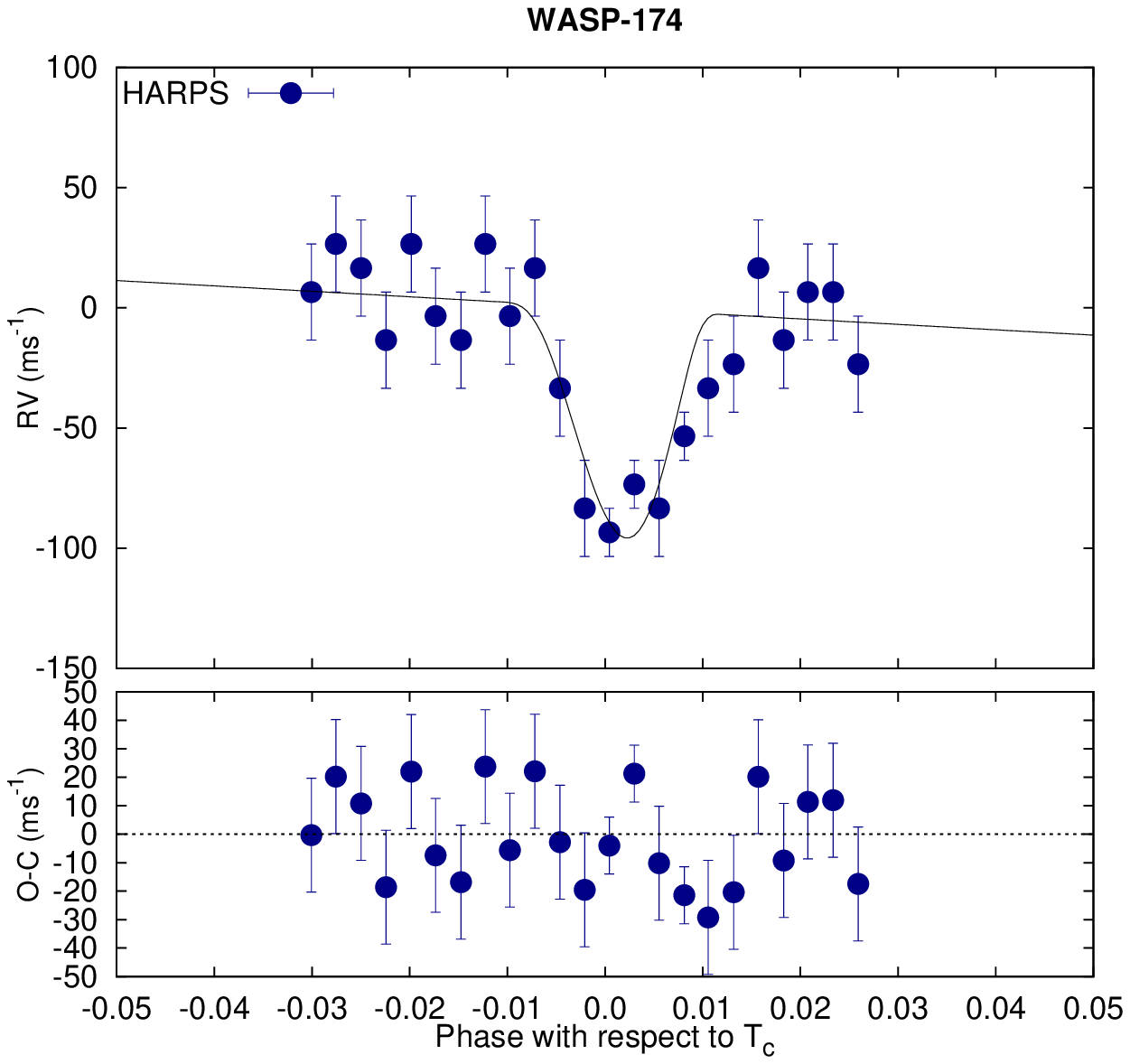}
\caption{A zoom of Figure~\ref{fig:RV} to highlight our fit of the 23 HARPS RVs, which were taken by the WASP team during a transit event \citep{temple:2018}, and the moderately misalignment of the orbit of WASP-174b.}
\label{fig:RMF}
\end{figure}

%-------------------------------------------------------------------
We acquired 16 RV measurements of WASP-174 by using the FEROS high-resolution spectrograph at the MPG~2.2\,m telescope. Comprehensive details about the instrument, data reduction, and corresponding RV extracting method can be found in the previous works of the HATSouth team (see footnote \#2), and are largely based on the CERES pipeline \citep{brahm:2017b}.  Details concerning the observations are summarized in Table~\ref{tab:specobs}, while the values of the RV measurements are reported in Table~\ref{tab:specmes} together with those from HARPS and Coralie taken from \citet{temple:2018}. The phase-folded RVs of WASP-174 are plotted in Figure~\ref{fig:RV} together with bisector spans (BS). The measurement of the Rossiter-McLaughlin effect by \citet{temple:2018} is highlighted in Figure~\ref{fig:RMF} (see discussion in Sect.~\ref{sec:discussion}).
\begin{table*}
\caption{Summary of spectroscopy observations.}
\label{tab:specobs}      
\centering                          
\begin{tabular}{lcccccc}        
\hline\hline   \\ [-6pt]               
~~~~~~~~Instrument & Date(s) & \# Spec. & Res. power & S/N Range\tablefootmark{a} & $\gamma_{\rm RV}$\tablefootmark{b} & RV Precision\tablefootmark{c} \\
 &  & & ($\rm {\lambda}$/$\rm{\Delta \lambda}$) & (km\,s$^{-1}$) & (km\,s$^{-1}$) \\
\hline        \\ [-6pt]                 
MPG~2.2\,m/FEROS & 2014 June -- 2015 April & 16 & 48000 & 41--114 & $4.9256 \pm 0.0068$ & 31 \\%
\hline                                   
\end{tabular}
\tablefoot{\\
\tablefoottext{a}{S/N per resolution element near 5180 \AA.}\\
\tablefoottext{b}{This is the zero-point RV from the best-fit orbit.}\\
\tablefoottext{c}{This is the scatter in the RV residuals from the best-fit orbit.}
}
\end{table*}  
%--------------------
%
\begin{table*}
\caption{Relative radial velocities and bisector spans for WASP-174.}
\label{tab:specmes}      
\centering                          
\begin{tabular}{lccrrccc}        
\hline\hline   \\ [-6pt] 
~~~~~~~~BJD  & RV\tablefootmark{a} & $\sigma_{\rm RV}$\tablefootmark{b} & BS~~~ & $\sigma_{\rm BS}$~~~ & Phase & Instrument & Reference \\
(2,450,000+) & (m\,s$^{-1}$) & (m\,s$^{-1}$) & (m\,s$^{-1}$) & (m\,s$^{-1}$) & \\
\hline \\ [-6pt] 
6823.67496 &  -84.77  & 34.00   & -35.00   &   11.00     &  0.299  &  FEROS    & This work \\
6836.57529 &  39.93   & 90.00   & 110.00   &   180.00    &  0.346  &  Coralie  & \citet{temple:2018} \\
6841.49461 &  -56.77  & 29.00   & -29.00   &   9.00      &   0.508  &  FEROS    & This work     \\
6842.49523 &  21.23   & 33.00   & -3.00    &   10.00     &  0.744  &  FEROS    & This work     \\
6845.50411 &  -3.77   & 29.00   & -20.00   &   9.00      &  0.455  &  FEROS    & This work     \\
6847.50175 &  6.23    & 26.00   & -21.00   &   9.00      &  0.927  &  FEROS    & This work     \\
6850.61870 &  15.23   & 46.00   & 19.00    &   13.00     &  0.663  &  FEROS    & This work     \\
6851.47603 &  33.23   & 29.00   & -79.00   &   9.00      &  0.865  &  FEROS    & This work     \\
6852.50485 &  -27.77  & 30.00   & 48.00    &   10.00     &  0.108  &  FEROS    & This work     \\
6852.61136 &  -5.77   & 33.00   & 18.00    &   10.00     &  0.133  &  FEROS    & This work     \\
6854.51539 &  35.23   & 37.00   & -36.00   &   11.00     &  0.583  &  FEROS    & This work     \\
6855.50546 &  59.23   & 29.00   & -3.00    &   9.00      &  0.817  &  FEROS    & This work     \\
6856.46921 &  -0.77   & 27.00   & 53.00    &   9.00      &  0.045  &  FEROS    & This work     \\
7072.73839 &  -90.07  & 60.00   & 170.00   &   120.00    &  0.127  &  Coralie  & \citet{temple:2018} \\
7225.57596 &  12.23   & 39.00   & 18.00    &   12.00     &  0.228  &  FEROS    & This work     \\
7233.57458 &  -16.77  & 36.00   & -20.00   &   11.00     &  0.117  &  FEROS    & This work     \\
7235.58577 &  67.23   & 58.00   & -144.00  &   15.00     &  0.592  &  FEROS    & This work     \\
7238.53027 &  15.23   & 35.00   & -42.00   &   11.00     &  0.287  &  FEROS    & This work     \\
7461.57183 &  -4.80   & 20.00   & -90.00   &   40.00     &  0.970  &  HARPS    & \citet{temple:2018} \\
7461.58250 &  15.20   & 20.00   & -160.00  &   40.00     &  0.972  &  HARPS    & \citet{temple:2018} \\
7461.59338 &  5.20    & 20.00   & -90.00   &   40.00     &  0.975  &  HARPS    & \citet{temple:2018} \\
7461.60425 &  -24.80  & 20.00   & -160.00  &   40.00     &  0.976  &  HARPS    & \citet{temple:2018} \\
7461.61514 &  15.20   & 20.00   & -150.00  &   40.00     &  0.980  &  HARPS    & \citet{temple:2018} \\
7461.62571 &  -14.80  & 20.00   & -100.00  &   40.00     &  0.983  &  HARPS    & \citet{temple:2018} \\
7461.63669 &  -24.80  & 20.00   & -120.00  &   40.00     &  0.985  &  HARPS    & \citet{temple:2018} \\
7461.64726 &  15.20   & 20.00   & -100.00  &   40.00     &  0.988  &  HARPS    & \citet{temple:2018} \\
7461.65792 &  -14.80  & 20.00   & -30.00   &   40.00     &  0.990  &  HARPS    & \citet{temple:2018} \\
7461.66870 &  5.20    & 20.00   & -140.00  &   40.00     &  0.993  &  HARPS    & \citet{temple:2018} \\
7461.67958 &  -44.80  & 20.00   & -90.00   &   40.00     &  0.995  &  HARPS    & \citet{temple:2018} \\
7461.69035 &  -94.80  & 20.00   & 120.00   &   40.00     &  0.998  &  HARPS    & \citet{temple:2018} \\
7461.70102 &  -104.80 & 10.00   & 70.00    &   20.00     &  0.000  &  HARPS    & \citet{temple:2018} \\
7461.71180 &  -84.80  & 10.00   & -30.00   &   20.00     &  0.003  &  HARPS    & \citet{temple:2018} \\
7461.72258 &  -94.80  & 20.00   & -130.00  &   40.00     &  0.005  &  HARPS    & \citet{temple:2018} \\
7461.73346 &  -64.80  & 10.00   & -150.00  &   20.00     &  0.008  &  HARPS    & \citet{temple:2018} \\
7461.74380 &  -44.80  & 20.00   & -100.00  &   40.00     &  0.010  &  HARPS    & \citet{temple:2018} \\
7461.75489 &  -34.80  & 20.00   & -0.00    &   40.00     &  0.013  &  HARPS    & \citet{temple:2018} \\
7461.76557 &  5.20    & 20.00   & -100.00  &   40.00     &  0.016  &  HARPS    & \citet{temple:2018} \\
7461.77655 &  -24.80  & 20.00   & -190.00  &   40.00     &  0.018  &  HARPS    & \citet{temple:2018} \\
7461.78701 &  -4.80   & 20.00   & -150.00  &   40.00     &  0.021  &  HARPS    & \citet{temple:2018} \\
7461.79799 &  -4.80   & 20.00   & -20.00   &   40.00     &  0.023  &  HARPS    & \citet{temple:2018} \\
7461.80887 &  -34.80  & 20.00   & -190.00  &   40.00     &  0.026  &  HARPS    & \citet{temple:2018} \\
7888.59786 &  -10.07  & 90.00   & -210.00  &   180.00    &  0.833  &  Coralie  & \citet{temple:2018} \\
7890.51593 &  -70.07  & 140.00  & 240.00   &   280.00    &  0.286  &  Coralie  & \citet{temple:2018} \\
7894.50253 &  -10.07  & 80.00   & -100.00  &   160.00    &  0.228  &  Coralie  & \citet{temple:2018} \\
7903.60566 &  -140.07 & 130.00  & -420.00  &   260.00    &  0.378  &  Coralie  & \citet{temple:2018} \\
7905.68911 &  49.93   & 70.00   & -300.00  &   140.00    &  0.870  &  Coralie  & \citet{temple:2018} \\
7917.56797 &  49.93   & 70.00   & -60.00   &   140.00    &  0.676  &  Coralie  & \citet{temple:2018} \\
7924.50566 &  19.93   & 60.00   & -350.00  &   120.00    &  0.315  &  Coralie  & \citet{temple:2018} \\
7951.50653 &  -70.07  & 170.00  & -410.00  &   340.00    &  0.692  &  Coralie  & \citet{temple:2018} \\
7954.49510 &  29.93   & 70.00   & 10.00    &   140.00    &  0.398  &  Coralie  & \citet{temple:2018} \\
7959.51703 &  -100.07 & 90.00   & -30.00   &   180.00    &  0.585  &  Coralie  & \citet{temple:2018} \\
7973.49257 &  19.93   & 120.00  & 170.00   &   240.00    &  0.886  &  Coralie  & \citet{temple:2018} \\
7974.51884 &  -150.07 & 90.00   & -160.00  &   180.00    &  0.128  &  Coralie  & \citet{temple:2018} \\
\hline                                 
\end{tabular}
\tablefoot{\\
\tablefoottext{a}{The zero-point of these velocities is arbitrary. An overall offset $\gamma_{\rm rel}$ fitted independently to the velocities from each instrument has been subtracted.}\\
\tablefoottext{b}{Internal errors excluding the component of astrophysical jitter considered in the joint analysis (Sect.~\ref{sec:analysis}).}
}
\end{table*}

%--------------------
%--------------------

%
%-------------------------------------------------------------------
\subsection{Photometric follow-up observations}
\label{sec:phot}
%-------------------------------------------------------------------
Broad-band photometric follow-up observations of several transit events of WASP-174b were also obtained with larger telescopes (see Table~\ref{tab:photobs_2}), with the aim to obtain a better determination of the photometric parameters. In particular, a complete transit was simultaneously observed with the AAT 3.9\,m and PEST~0.3\,m telescopes and another complete transit was recorded with the LCOGT~1\,m. 

\begin{figure}
\centering
\includegraphics[width=\hsize]{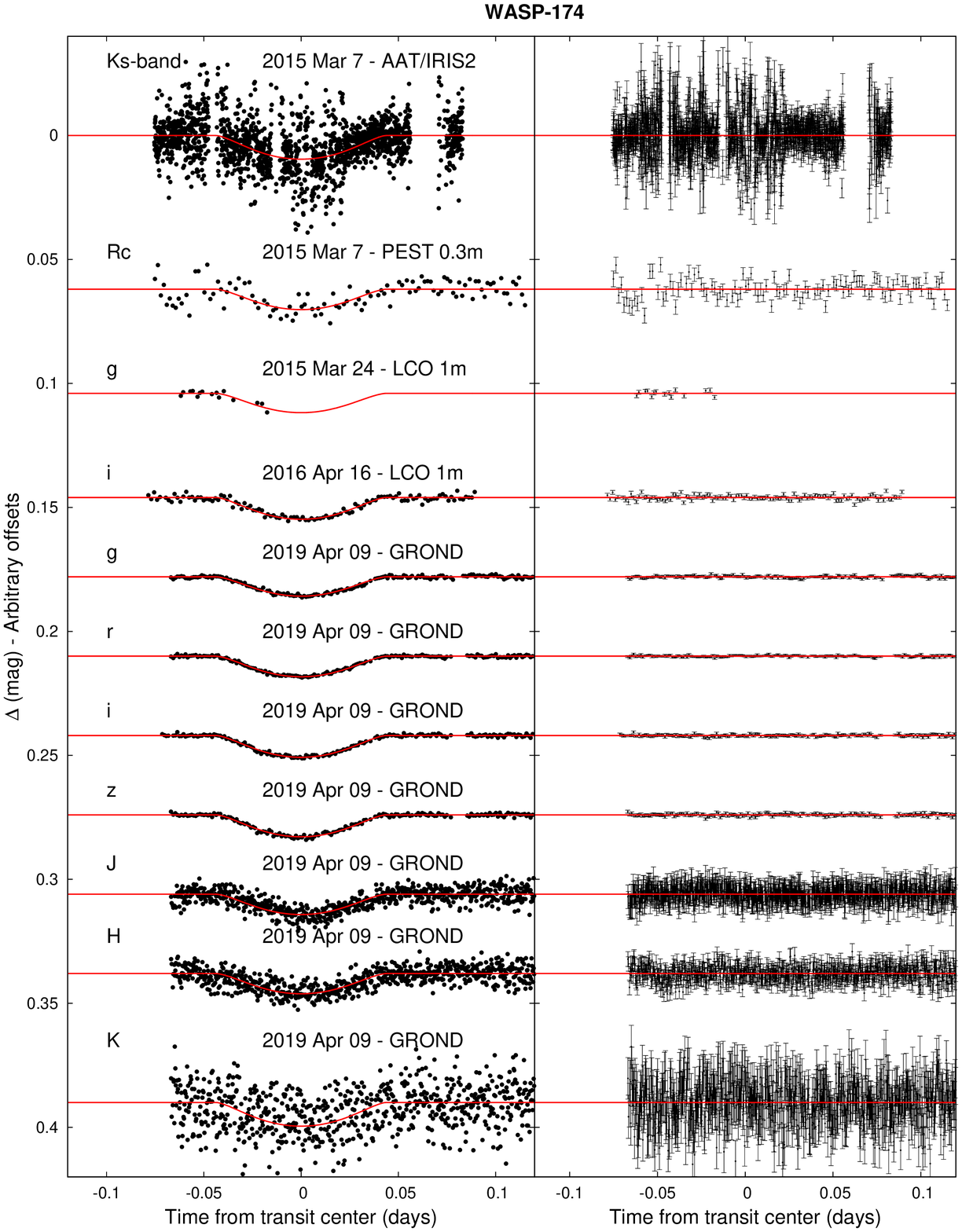}
\caption{Unbinned transit light curves for WASP-174. The dates of the events, filters and instruments used are indicated. Light curves following the first are displaced vertically for clarity. Our best fit from the global modeling is shown by the solid lines. The residuals from the best-fitting model are shown on the right-hand-side in the same order as the original light curves. Note the precision of the GROND light curves, which have point-to-point scatters between 0.37~mmag and 0.52~mmag.}
\label{fig:lcs}
\end{figure}

More recently, we obtained high-quality, optical light curves of another transit event, by using the GROND multi-band camera mounted on the MPG 2.2\,m telescope at the ESO Observatory in La Silla, during a photometric night with good seeing ($\approx 0.5^{\prime \prime}$). GROND is an instrument which is capable of simultaneous observations in four optical (similar to Sloan $g^{\prime}$, $r^{\prime}$, $i^{\prime}$, $z^{\prime}$) and three NIR ($J$, $H$, $K$) passbands \citep{greiner:2008}. It was conceived for observing transient events, but it turned out to be an optimum instrument for monitoring planetary transits too \citep[e.g.][]{mohler:2013}. Several points were cut from the final light curves due to tracking issues before and after the transit. Again, details about the telescopes and data reduction are reported in the previous works of the HATSouth team \citep[for the GROND-data reduction, see also][]{mancini:2019}. The corresponding light curves can be examined in Figure~\ref{fig:lcs}.

\begin{table*}
\caption{Summary of photometric follow-up observations.}
\label{tab:photobs_2}      
\centering                          
\begin{tabular}{lcrrcc}        
\hline\hline                 
Telescope/Instrument & Date(s) & \# Images & Cadence & Filter & Precision \\
 &  & & (s)~~~ &  & (mmag) \\
\hline                        
~~~~AAT 3.9\,m/IRIS2 & 2015 Mar 07 & 1758~~ & 6~~~ & $K_{\rm {s}}$ & 8.26 \\
~~~~PEST~0.3\,m      & 2015 Mar 07 & 198~~ & 132~~~ & $R_{\rm {c}}$ & 3.68 \\
~~~~LCOGT~1\,m/SBIG  & 2015 Mar 24 & 15~~ & 212~~~ & $g$ & 1.08 \\
~~~~LCOGT~1\,m/sinistro & 2016 Apr 16 & 112~~ & 129~~~ & $i$ & 1.00 \\
~~~~MPG~2.2\,m/GROND & 2019 Apr 10 & 177~~ & 90~~~  & $g^{\prime}$  & 0.47 \\
~~~~MPG~2.2\,m/GROND & 2019 Apr 10 & 175~~ & 90~~~  & $r^{\prime}$  & 0.37 \\
~~~~MPG~2.2\,m/GROND & 2019 Apr 10 & 178~~ & 90~~~  & $i^{\prime}$  & 0.44 \\
~~~~MPG~2.2\,m/GROND & 2019 Apr 10 & 175~~ & 90~~~  & $z^{\prime}$  & 0.52 \\
~~~~MPG~2.2\,m/GROND & 2019 Apr 10 & 741~~ & 22~~~  & $J$  & 2.67 \\
~~~~MPG~2.2\,m/GROND & 2019 Apr 10 & 741~~ & 22~~~  & $H$  & 3.10 \\
~~~~MPG~2.2\,m/GROND & 2019 Apr 10 & 741~~ & 22~~~  & $K$  & 8.30 \\
~~~~TESS & 2019 Mar 29 -- Apr 04 & 975~~ & 1800~~~  & TESS filter & 1.17 \\
\hline                                   
\end{tabular}
\end{table*}  

%
%-------------------------------------------------------------------
\subsection{Limb darkening and transit depth}
\label{sec:phot}
\begin{figure}
\centering
\includegraphics[width=\hsize]{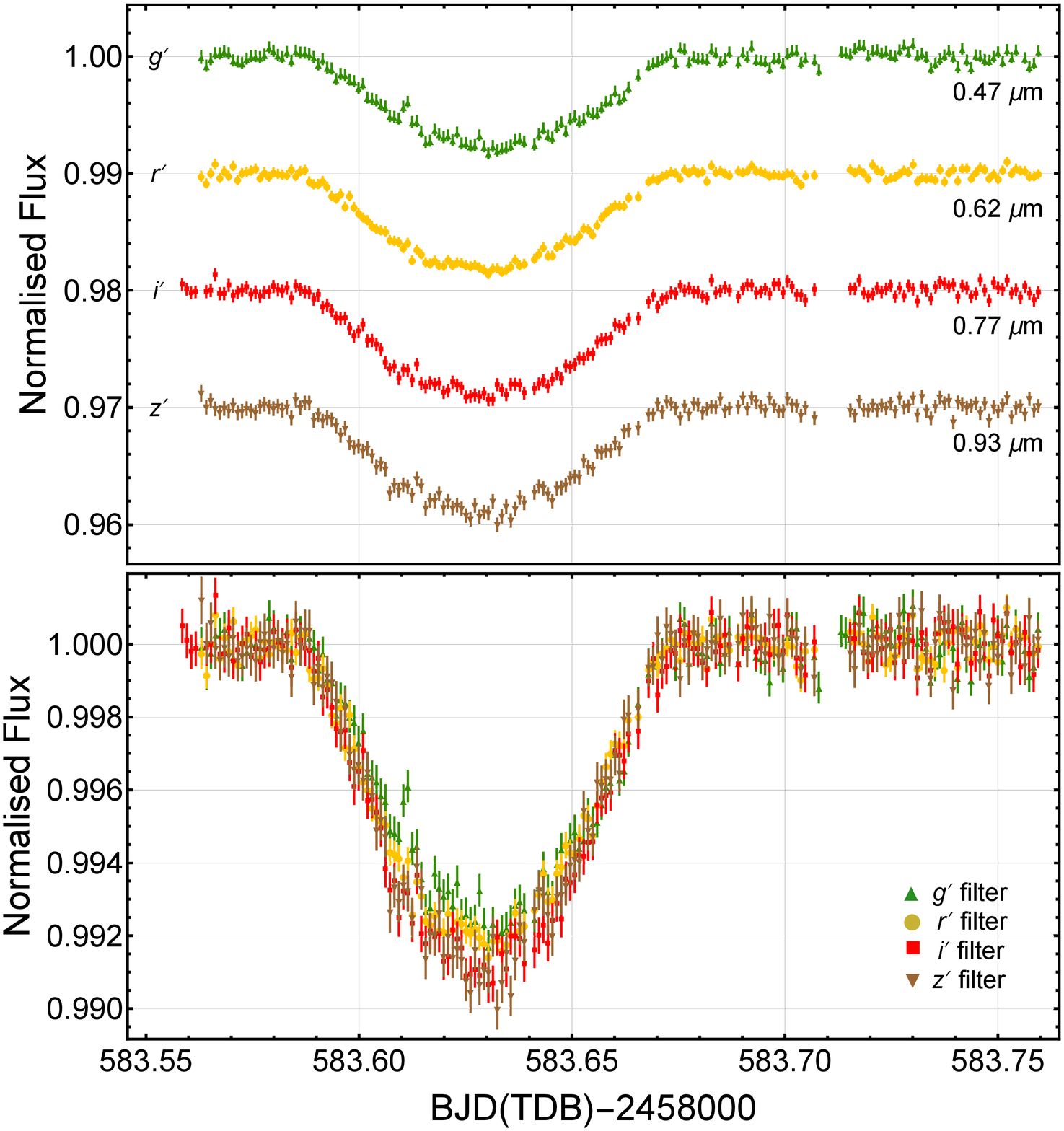}
\caption{The simultaneous four optical light curves of a transit of WASP-174b obtained with GROND. {\it Top panel}: Individual light-curves with arbitrary offsets for clarity; {\it Bottom panel}: All light-curves superimposed for comparison purposes. The passbands and their central wavelengths are labelled. Note how the transit light curve shape changes with wavelength. Contrary to higher-inclination transiting planetary systems, the transit in the $g^{\prime}$ band is shallower than the other bands, because the transit is grazing and the limb darkening is stronger at bluer wavelengths. The same effect was observed in only another case, that is WASP-67b \citep{mancini:2014}.}
\label{fig:grond-lc}
\end{figure}

%-------------------------------------------------------------------
We plotted the four optical GROND light curves superimposed in Figure~\ref{fig:grond-lc} to highlight the differences of the transit depth among the four passbands. For normal transiting-planet systems, in which the orbital inclination is $\approx 90^{\circ}$ and it is possible to precisely measure the timings of the four contact points, it is well known that the transit depth gradually decreases moving from blue to red bands \citep[e.g.][]{knutson:2007};  actually, the limb radiation of a star is characteristically cooler than that from its centre and, hence, is slightly redder. Here, instead, we observe the opposite effect. Since WASP-174b is a grazing transiting planet, during its transits it only covers the limb of its parent start, which is fainter in blue than in red wavelengths, resulting in shallower eclipses in the bluest bands for this system. This behavior is similar to that of WASP-67b, another grazing transiting planet \citep{mancini:2014} \citep[see also][pag.~224]{perryman:2018}.

% TESS
%-------------------------------------------------------------------
\subsection{TESS photometry}
\label{sec:tess}
%-------------------------------------------------------------------
WASP-174 was also observed by TESS \citep{ricker:2015} during Sector 10 of its primary mission. Near continuous observations of the target star from 2019-03-29 to 2019-04-22 were obtained through Camera 2, CCD 3 of the spacecraft, and downlinked via the $\sim 30$ minute cadenced Full Frame Images (FFI).

The light curve was extracted using the calibrated FFIs from the TESS Science Processing Operations Center \citep{jenkins:2016}. 
A cut out of $10\times10$ pixels around the target star was downloaded via the \emph{TESScut} tool from the Mikulski Archive for Space Telescopes (MAST) archives. Aperture photometry was performed using the \emph{lightkurve} package \citep{barentsen:2019}, with the aperture manually selected to best avoid the $T_\mathrm{mag} = 12.7$ variable star $24.5 \arcsec$ away from WASP-174. The light curve was detrended via a set of cubic splines, fitted with the known transits and spacecraft momentum dumps masked out. Finally, the light curve was adjusted for third light blending due to nearby stars via the TIC-6 catalogue \citep{stassun:2018}. The last step was necessary, because this nearby $12.7$-mag star is somewhat variable, causing contamination on the WASP-174 star. While no clear hints of occultation are present, a very clear transit signal was obtained (Figure~\ref{fig:tess_fit}), which is in close agreement to the transits presented in the previous sections.

% Analysis
%-------------------------------------------------------------------
\section{Physical properties}
\label{sec:analysis}
\begin{figure*}
\centering
\includegraphics[width=\hsize]{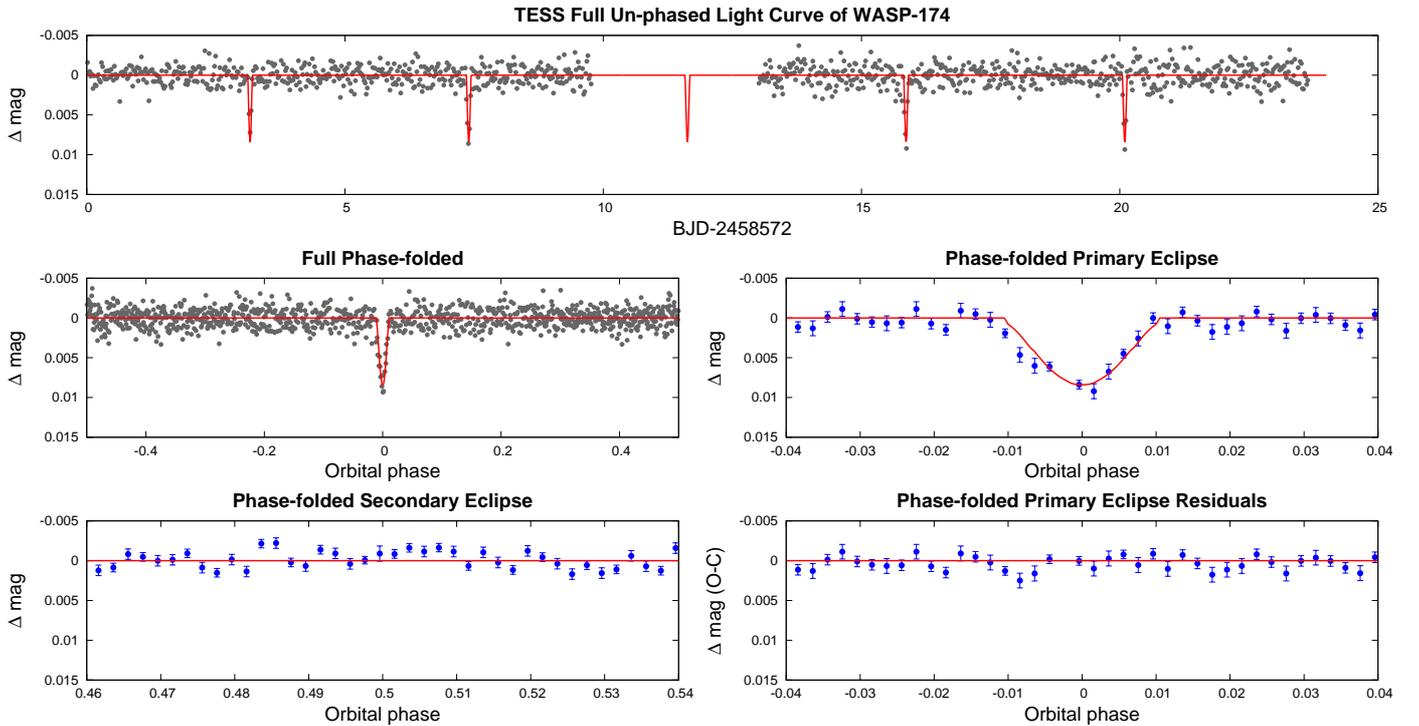}
\caption{TESS light curve of WASP-174 and our best fit from the global modeling (solid lines). The full un-phased light curve is plotted in the top panel. The left-hand middle panel shows the full phase-folded light curve, while the phase-folded light curve zoomed-in on the secondary eclipse is displayed on the left-hand bottom panel. The right-hand middle panel shows the phase-folded light curve zoomed-in on the transit, while the corresponding residuals from the best-fitting model are displayed in the right-hand bottom panel.}
\label{fig:tess_fit}
\end{figure*}

%-------------------------------------------------------------------
The observational data, which were presented in the previous section,
were modeled to refine the main physical properties of the WASP-174
system. The modeling was performed following the procedure described
in detail in \citet{hartman:2019} with a few modifications that we
discuss here. We carry out a joint fit of the available light curves,
RV observations, broad-band catalog photometry, and spectroscopically
determined stellar atmospheric parameters using a Differential
Evolution Markov Chain Monte Carlo (DEMCMC) procedure. We constrain
the physical parameters of the star (and planet) using two different
methods: (1) an isochrone-based method where we use the PARSEC stellar
evolution models \citep{marigo:2017} and perform an interpolation
within a pre-computed grid of these isochrones at each step in the
Markov Chain using the stellar effective temperature, density and
metallicity (which are varied in the fit) as look-up parameters; (2)
an empirical model where we use the stellar radius (determined from
the temperature and luminosity, which is constrained primarily by the
Gaia observations) and density to empirically determine the stellar
mass. In this paper we differ from \citet{hartman:2019} in using a
newer version of the PARSEC models obtained through the CMD v3.2
web-interface\footnote{\url{http://stev.oapd.inaf.it/cgi-bin/cmd}},
adopting an updated set of bolometric corrections for the Gaia DR2
photometric bandpasses \citep{maizapellaniz:2018}, and incorporating
infrared photometry from the WISE mission \citep{cutri:2014} into the
fit as well. Incorporating the WISE photometry allows for a more
direct constraint on the extinction $A_{V}$, whereas this was
previously constrained primarily through the MWDUST 3D dust model of
the Galaxy \citep{bovy:2016}. We still use the MWDUST model as a
prior, but it is less informative on the value than in our previous
analyses. To model the Rossiter-McLaughlin effect observed in the
HARPS measurements presented by \citet{temple:2018}, which we fit in
the joint analysis (see Figure~\ref{fig:RMF}), we use the ARoME package \citep{boue:2013},
utilizing the modeling branch appropriate for RV observations obtained
through the cross-correlation function (CCF) method. 
In fitting the TESS light curve, the model is integrated over the 30 minute full-frame image exposure time. We also fit the detrended TESS light curve, rather than detrending simultaneously with the fit, see Figure~\ref{fig:tess_fit}.

One other difference from \citet{hartman:2019} is that we allow the quadratic limb darkening coefficients to vary in the fit, whereas previously we held them fixed to theoretically expected values. Here we allow them to vary, but use the tabulations from \citet{claret:2012,claret:2013} and \citet{claret:2018} to place Gaussian priors on their values.

\begin{table*}
\caption{Derived Stellar Parameters for WASP-174. Both the isochrone- and empirical-model results are reported and compared with those from the discovery paper.}
\label{tab:stellarparameters2}      
\centering  
\begin{tabular}{lccc}
\hline
Quantity & \multicolumn{2}{c}{This work} &\citet{temple:2018}\\
\hline  \\[-6pt]%%
& \textbf{Isochrone} & \textbf{Empirical} &  \\
& (fiducial) &  &  \\ [2pt]
~~~~\Teff$_{\star}$ (K)\dotfill & {\ensuremath{6399 \pm 68}} & {\ensuremath{6376\pm59}} & $6400 \pm 100$ \\
~~~~[Fe/H]\dotfill & {\ensuremath{0.060\pm0.060}} & {\ensuremath{0.013\pm0.073}}  & $0.09 \pm 0.09$ \\
~~~~$v \sin{i}$ (\kms)\dotfill & {\ensuremath{16.24 \pm 0.53}} & {\ensuremath{16.35 \pm 0.51}}  & $16.5 \pm 0.5$ \\
~~~~$M_{\star}$ (\Msun) \dotfill & {\ensuremath{1.240 \pm 0.038}} & {\ensuremath{1.303 \pm 0.090}}  & $1.27 \pm 0.06$ \\ [2pt]
~~~~$R_{\star}$ (\Rsun) \dotfill & {\ensuremath{1.347 \pm 0.018}} & {\ensuremath{1.365 \pm 0.032}}  & $1.3 \pm 0.1$ \\ [2pt]
~~~~$\logg_{\star}$ (cgs) \dotfill & {\ensuremath{4.273 \pm 0.010}} & {\ensuremath{4.281 \pm 0.020}}  & $4.35 \pm 0.05$ \\ [2pt]
~~~~$\rho_{\star}$ (g\,cm$^{-3}$) \dotfill & {\ensuremath{0.716 \pm 0.022}} & {\ensuremath{0.715_{-0.028}^{+0.037}}}  & $0.85 \pm 0.28$ \\ [2pt]
~~~~$L_{\star}$ ($L_{\odot}$) \dotfill & {\ensuremath{2.73 \pm 0.16}} & {\ensuremath{2.77 \pm 0.15}}  & -- \\  [2pt]
~~~~Age (Gyr)\dotfill & {\ensuremath{2.20 \pm 0.52}} & {\ensuremath{1.46_{-0.80}^{+1.84}}}  & $1.65 \pm 0.85$ \\ [2pt]
~~~~$A_{V}$ (mag)\dotfill & {\ensuremath{0.114 \pm 0.044}} & {\ensuremath{0.079_{-0.028}^{+0.038}}}  & -- \\ [2pt]
~~~~Distance (pc)\dotfill & {\ensuremath{406.8 \pm 6.6}} & {\ensuremath{413.2 \pm 9.0}}  & -- \\[2pt]
\hline
\end{tabular}
%\tablefoot{\\}
\end{table*}
\begin{table*}
\caption{Orbital and planetary parameters for WASP-174b. Both the isochrone- and empirical-model results are reported and compared with those from the discovery paper.}
\label{tab:planetaryparameters}      
\centering  
\begin{tabular}{lccc}
\hline
Parameter & \multicolumn{2}{c}{This work} &\citet{temple:2018}\\
\hline  \\[-6pt]%%
& \textbf{Isochrone} & \textbf{Empirical} &  \\
& (fiducial) &  &  \\ [2pt]
\multicolumn{1}{l}{\textbf{Light curve parameters}} \\
~~~$P$ (days) \dotfill & {\ensuremath{4.2337005 \pm 0.0000018}} & {\ensuremath{4.2337006 \pm 0.0000017}} & $4.233700 \pm 0.000003$ \\
~~~$T_c$ (${\rm BJD}$) \dotfill & {\ensuremath{2458545.52660 \pm 0.00014}} & {\ensuremath{2458545.52661 \pm 0.00013}} & $2457465.9322 \pm 0.0005$ \\
~~~$T_{14}$ (days) \dotfill & {\ensuremath{0.08726 \pm 0.00065}} & {\ensuremath{0.08739 \pm 0.00073}} & $0.084 \pm 0.002$ \\
~~~$a/R_{\star}$ \dotfill & {\ensuremath{8.789 \pm 0.088}} & {\ensuremath{8.78_{-0.12}^{+0.15}}} & -- \\[2pt]
~~~$\zeta/R_{\star}$ \dotfill & {\ensuremath{43.1 \pm 2.2}} & {\ensuremath{42.6_{-2.8}^{+4.8}}} & -- \\
~~~$R_{p}/R_{\star}$ \dotfill & {\ensuremath{0.1098 \pm 0.0030}} & {\ensuremath{0.1100 \pm 0.0044}} & -- \\ [2pt]
~~~$R^2_{p}/R^2_{\star}$ \dotfill & -- & -- & $0.0089 \pm 0.0003$ \\ [2pt]
~~~$b^2$ \dotfill & {\ensuremath{0.908_{-0.013}^{+0.011}}} & {\ensuremath{0.907_{-0.018}^{+0.019}}} & -- \\ [2pt]
~~~$b \equiv a \cos{i}/R_{\star}$ \dotfill & {\ensuremath{0.9530_{-0.0066}^{+0.0060}}} & {\ensuremath{0.9523_{-0.0095}^{+0.0100}}} & $0.92 \pm 0.03$ \\ [2pt]
~~~$i$ (deg) \dotfill & {\ensuremath{83.780 \pm 0.096}} & {\ensuremath{83.77 \pm 0.15}} & $84.5 \pm 0.6$ \\  [4pt]
\multicolumn{1}{l}{\textbf{Limb-darkening coefficients\,\tablefootmark{a}}} \\
~~~$c_1,g$ \dotfill & {\ensuremath{0.39 \pm 0.13 }}   & {\ensuremath{0.40 \pm 0.13}} & -- \\
~~~$c_2,g$ \dotfill & {\ensuremath{0.21\pm0.15}}   & {\ensuremath{0.22\pm0.15}} & -- \\
~~~$c_1,r$ \dotfill & {\ensuremath{0.30\pm0.12}}   & {\ensuremath{0.29\pm0.12}} & -- \\
~~~$c_2,r$ \dotfill & {\ensuremath{0.18\pm0.14}}   & {\ensuremath{0.21\pm0.14}} & -- \\
~~~$c_1,R$ \dotfill & {\ensuremath{0.24\pm0.15}}   & {\ensuremath{0.22\pm0.14}} & -- \\
~~~$c_2,R$ \dotfill & {\ensuremath{0.20\pm0.17}}   & {\ensuremath{0.25\pm0.18}} & -- \\[2pt]
~~~$c_1,i$ \dotfill & {\ensuremath{0.18\pm0.11}} & {\ensuremath{0.20 \pm 0.10}} & -- \\[2pt]
~~~$c_2,i$ \dotfill & {\ensuremath{0.13\pm0.14}}   & {\ensuremath{0.14\pm0.12}} & -- \\
~~~$c_1,z$ \dotfill & {\ensuremath{0.129 \pm 0.089}} & {\ensuremath{0.141 \pm 0.091}} & -- \\
~~~$c_2,z$ \dotfill & {\ensuremath{0.20\pm0.12}}   & {\ensuremath{0.19 \pm 0.12}} & -- \\[2pt]
~~~$c_1,J$ \dotfill & {\ensuremath{0.26 \pm 0.13 }} & {\ensuremath{0.29 \pm 0.13}} & -- \\ [2pt]
~~~$c_2,J$ \dotfill & {\ensuremath{0.30 \pm 0.15}} & {\ensuremath{0.28 \pm 0.16}} & -- \\  [2pt]
~~~$c_1,H$ \dotfill & {\ensuremath{0.35 \pm 0.13}} & {\ensuremath{0.32 \pm 0.13 }} & -- \\ [2pt]
~~~$c_2,H$ \dotfill & {\ensuremath{0.23 \pm 0.16}} & {\ensuremath{0.29 \pm 0.16}} & -- \\  [2pt]
~~~$c_1,K$ \dotfill & {\ensuremath{0.142_{-0.099 +0.142}}} & {\ensuremath{0.16 \pm 0.11}} & -- \\ [2pt]
~~~$c_2,K$ \dotfill & {\ensuremath{0.19 \pm 0.16}} & {\ensuremath{0.18 \pm 0.15}} & -- \\  [2pt]
~~~$c_1,{\rm TESS}$ \dotfill & {\ensuremath{0.19 \pm 0.12}} & {\ensuremath{0.17 \pm 0.11}} & -- \\ [2pt]
~~~$c_2,{\rm TESS}$ \dotfill & {\ensuremath{0.10 \pm 0.14}} & {\ensuremath{0.11 \pm 0.15}} & -- \\  [4pt]
\multicolumn{1}{l}{\textbf{RV properties}} \\
~~~$K$ (\ms) \dotfill & {\ensuremath{36.1\pm 9.8}} & {\ensuremath{38.4\pm 9.1}} & -- \\
~~~$e$ \dotfill & {\ensuremath{0}} & {\ensuremath{0}}  & $0$ \\
~~~RV jitter FEROS (\ms)   \dotfill & {\ensuremath{<6.0}} & {\ensuremath{<6.1}}  & -- \\
~~~RV jitter HARPS (\ms)   \dotfill & {\ensuremath{<2.8}} & {\ensuremath{<12.6}} & -- \\
~~~RV jitter CORALIE (\ms) \dotfill & {\ensuremath{<19.2}} & {\ensuremath{<25.7}} & -- \\   [4pt]
\multicolumn{1}{l}{\textbf{Planetary parameters}} \\ 
~~~$M_{\rm p}$ ($M_{\rm Jup}$) \dotfill & {\ensuremath{0.330 \pm 0.091}} & {\ensuremath{0.366 \pm 0.088}} & $<1.3$ \\ [2pt]
~~~$R_{\rm p}$ ($R_{\rm Jup}$) \dotfill & {\ensuremath{1.437 \pm 0.050}} & {\ensuremath{1.462 \pm 0.075}} & $1.2 \pm 0.5$ \\ [2pt]
~~~$C(M_{\rm p},R_{\rm p})$  \dotfill & {\ensuremath{-0.00}} & {\ensuremath{0.15}} & -- \\ [2pt]
~~~$\rho_{\rm p}$ (g\,cm$^{-3}$) \dotfill & {\ensuremath{0.135 \pm 0.042}} & {\ensuremath{0.143_{-0.033}^{+0.047}}} & -- \\ [2pt]
~~~$\log{g_{\rm p}}$ (cgs)       \dotfill & {\ensuremath{2.59 \pm 0.13}} & {\ensuremath{2.63 \pm 0.12}} & -- \\ [2pt]
~~~$a$ (au) \dotfill & {\ensuremath{0.05503 \pm 0.00056}} & {\ensuremath{0.0560_{-0.0016}^{+0.0012}}} & $0.0555 \pm 0.0009$ \\ [2pt]
~~~$T_{\rm eq}$ (K) \dotfill & {\ensuremath{1528 \pm 17 }} & {\ensuremath{1521 \pm 19}} & $1557\pm14$ \\ [2pt]
~~~$\Theta$ \dotfill & {\ensuremath{0.0203 \pm 0.0056}} & {\ensuremath{0.0214 \pm 0.0051}} & -- \\ [2pt]
~~~$\log_{10}\langle F \rangle$ (cgs) \dotfill & {\ensuremath{9.089 \pm 0.019}} & {\ensuremath{9.085 \pm 0.021}} & -- \\ [2pt]
\hline
\end{tabular}
\tablefoot{\\
\tablefoottext{a}{A quadratic law was used and the LD coefficients were allowed to vary, but with Gaussian priors adopted from the tabulations by \citet{claret:2012,claret:2013} and \citet{claret:2018}.}\\}
\end{table*}

\begin{figure}
\centering
\includegraphics[width=\hsize]{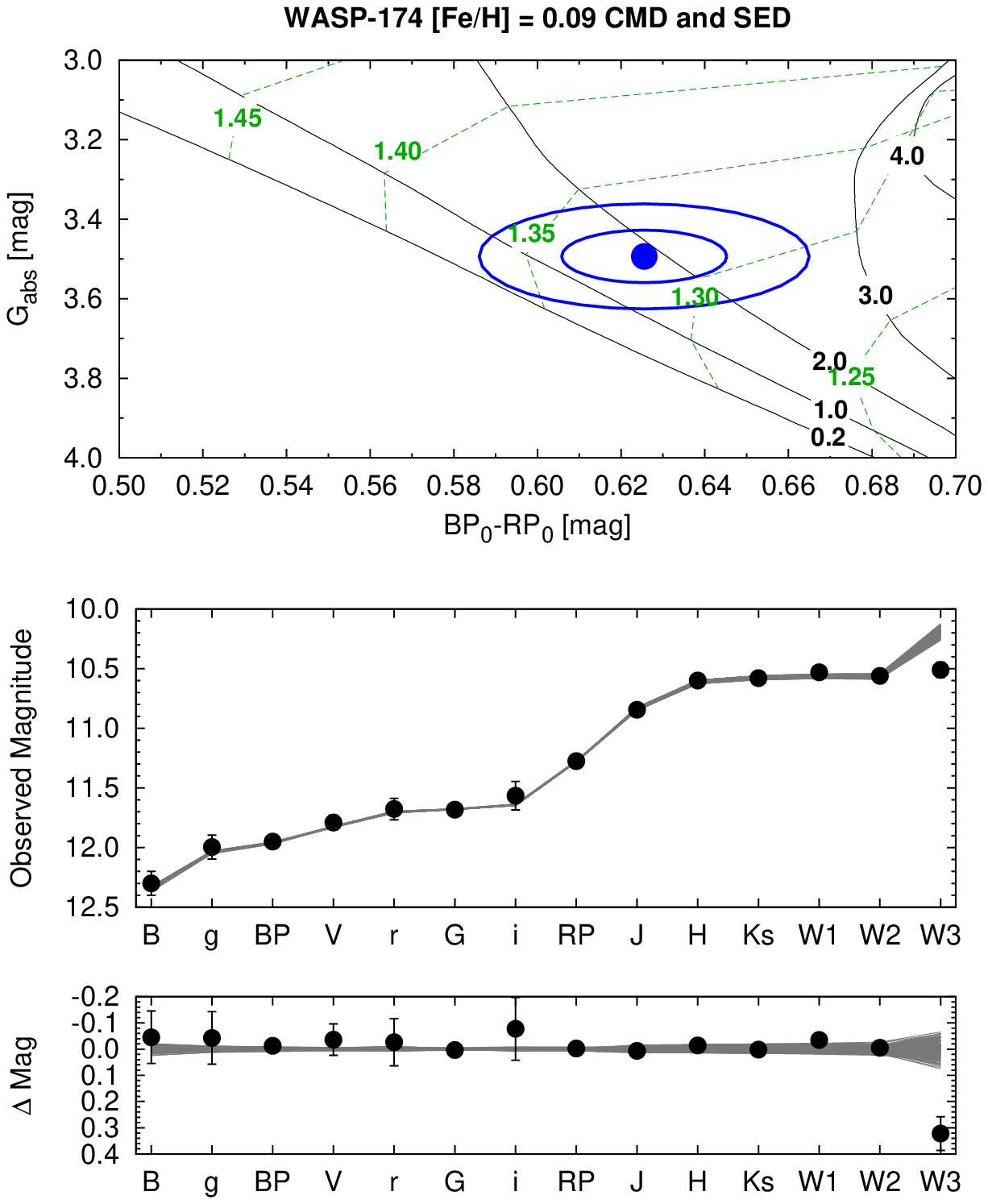}
\caption{{\it Top panel:} Color-magnitude diagram of WASP-174 compared to theoretical isochrones (black lines) and stellar evolution tracks (green lines) from the PARSEC models interpolated at its metallicity, which was spectroscopically determined. The stellar mass of each evolution track is reported in green in solar mass units, while the black numbers denote the age of of each isochrone in Gyr. The blue circle shows the measured reddening- and distance-corrected value from Gaia DR2, together with $1\sigma$ and $2\sigma$ confidence regions (blue lines). {\it Bottom panels:} Broadband spectral energy distribution (black points) as photometrically measured through the observed filters, compared with model SEDs (gray lines), followed by the O-C residuals from the best-fit model.}
\label{fig:CMD_SED}
\end{figure}

The stellar parameters for the star and the planet are shown in Table~\ref{tab:stellarparameters2} and \ref{tab:planetaryparameters}, respectively. We reported both the isochrone- and empirical-model results and compared them with those from the discovery paper. 
The results from the two models are consistent with each other to within $2\,\sigma$, with the isochrone models being somewhat more precise. We adopt the sets of parameters coming from the isochrone model as the final ones for consistency with our previous works, for which the isochrone was always our default fiducial model. The location of WASP-174 on the absolute $G$-magnitude versus Gaia DR2 $BP-RP$ color in illustrated in Figure~\ref{fig:CMD_SED} (top panel), while the broadband spectral energy distribution (SED) fits to the observed bands is plotted in the bottom panel of the same figure.

Our analysis gives a grazing solution with a confidence greater than $5\sigma$, that is $b + (R_{\rm p}/R_{\star}) =1.0628 \pm 0.0094$, which is larger than 1 and, therefore, indicates that the transits of WASP-174b are grazing. 
Using plane geometry and circle-circle intersection formulae \citep{lillo-box:2015}, we estimate that roughly $76\%$ of the planetary disc transits the parent star at each mid-transit time. Specifically, $A_{\rm ecl}/A_{\rm p}=0.76 \pm 0.12$, where $A_{\rm p}$ is the full area of the disk of the planet, while $A_{\rm ecl}$ is the area of the part of the planetary disk that actually eclipses the parent star.

With our new data, we are able to constrain the mass and radius of the planet. Based on the results from the isochrone model, we find that WASP-174b has $M_{\rm p}=0.330\pm0.091 \, M_{\rm Jup}$ and $R_{\rm p}=1.437\pm0.050 \, R_{\rm Jup}$, and is therefore a very low-density, inflated hot giant planet with $T_{\rm eq}=1528 \pm 17$\,K. It is worth noting that the empirical model points towards a planet with greater gravity and density, even though the differences versus the fiducial parameters are within the error bars. 

% Discussion
%-------------------------------------------------------------------
\section{Summary and discussion}
\label{sec:discussion}
\begin{figure}
\centering
\includegraphics[width=\columnwidth]{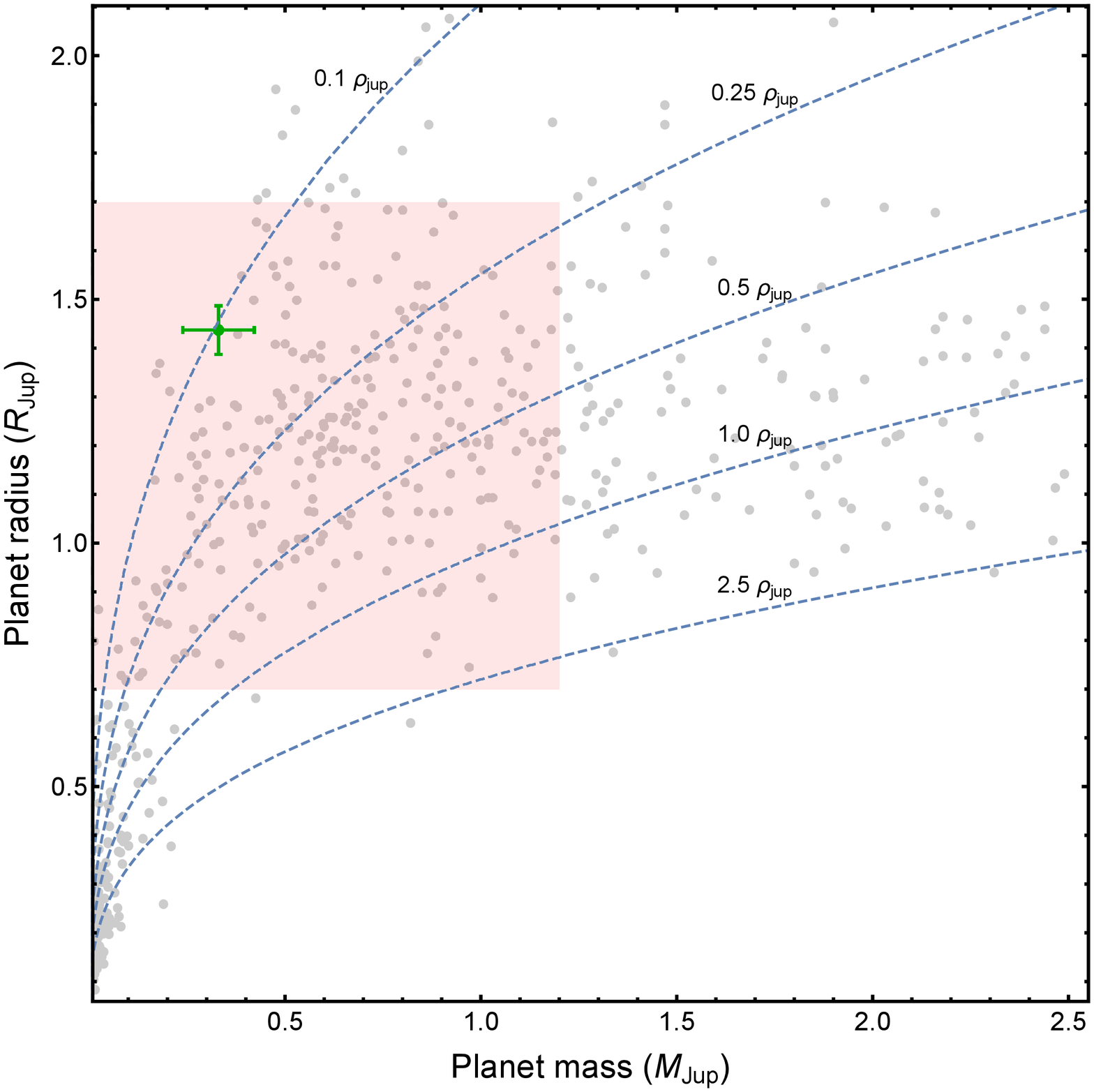}
\caption{
The masses and radii of the known transiting extrasolar planets. The plot is restricted to exoplanets with values of the mass until $2.5\,M_{\rm Jup}$ and radius until $2.0\,R_{\rm Jup}$. Grey points denote values taken from the Transiting Extrasolar Planet Catalogue (TEPCat), which is available at \texttt{http://www.astro.keele.ac.uk/jkt/tepcat/} \citep{southworth:2011}. Their error bars have been suppressed for clarity. The green point with error bars refers to the position of WASP-174b in the diagram (this work), compared to the possible values of the two parameters reported in the discovery paper (light-red box; \citealp{temple:2018}). Dotted lines show where density is 2.5, 1.0, 0.5, 0.25 and 0.1 $\rho_{\rm Jup}$.
}
\label{fig:mass-radius}
\end{figure}

%-------------------------------------------------------------------
%
In this paper we have refined the physical parameters of the transiting planetary system WASP-174, which was a HATSouth transiting planet candidate at the time its discovery was announced by the WASP team in \citet{temple:2018}. In particular, using the data from the HATSouth survey and those from spectroscopic and photometric follow-up observations, we were able to confirm that the WASP-174b is transiting in front of its parent star every $\approx 4.2$\,days and the semi-major axis of its orbit is $\approx 0.055$\,au. We also confirm the main characteristics of the host star, a F6\,V dwarf star with $M_{\star}=1.240 \pm 0.038 \, M_{\sun}$ and $R_{\star}=1.347\pm0.018 \, R_{\sun}$ (Table~\ref{tab:stellarparameters2}). Most importantly, we were able to obtain a much better characterization of the physical parameters of the exoplanet WASP-174b with respect to the values reported in \citet{temple:2018}, see Table~\ref{tab:planetaryparameters} and Figure~\ref{fig:mass-radius}.

Moreover, while \citet{temple:2018} were not able to precisely distinguish between a grazing or near-grazing solution, we have found that $b+(R_{\rm p}/R_{\star})>1$ at more than $5\sigma$ confidence level, which definitively points towards a grazing nature for the WASP-174b transits. We have estimated that at mid-transit roughly $76\%$ of the planetary disk eclipses the host star.
The grazing scenario is also suggested by observing the shape and the depth of the grazing transits of WASP-174b (see Figure~\ref{fig:grond-lc}) that change with the wavelength in a reverse way with respect to normal cases of a planetary transit, where the disk of the planet is entirely inside the disk of its parent star. Here, instead, the transits are grazing and the planet covers only the limb of the star, causing an increasing of the transit depth moving from blue to red bands. 

\begin{figure}
\centering
\includegraphics[width=\columnwidth]{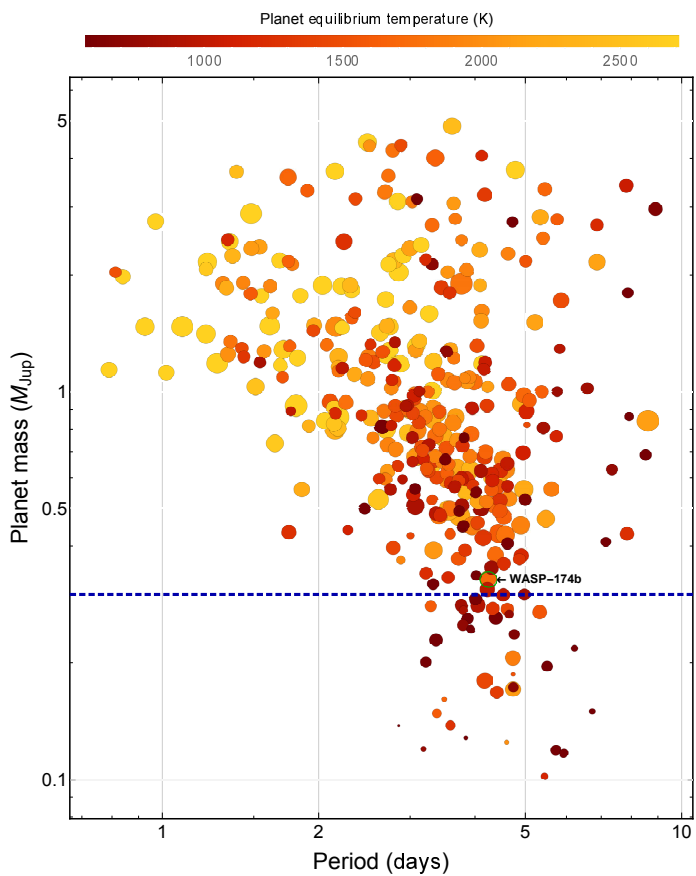}
\caption{Mass-period diagram of all known transiting hot giants, i.e. transiting exoplanets in the mass range $0.1\,M_{\rm Jup} < M_{\rm p} < 5\,M_{\rm Jup}$ and with an orbital period $\leq10$\,days. The planets are represented by circles, whose size is proportional to their radius. Color indicates equilibrium temperature. The error bars have been suppressed for clarity. The positions of WASP-174b is highlighted. Data taken from TEPCat. The horizontal dashed line indicates the mass limit of Jupiter planets ($0.3\,M_{\rm Jup}$) according to the definition given by \citet{hatzes:2015}.}
\label{fig:mass-period}
\end{figure}

\begin{figure}
\centering
\includegraphics[width=\columnwidth]{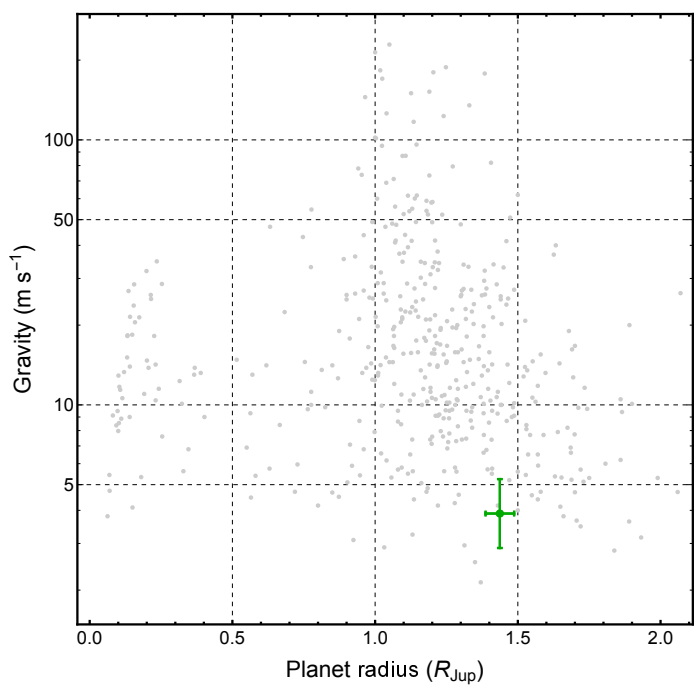}
\caption{
The radius-gravity diagram of the currently known transiting exoplanets with precisely measured mass and radius. The grey points denote values taken from TEPCat. Their error bars have been suppressed for clarity. The position of WASP-174b is shown in green with error bars.}
\label{fig:radius-gravity}
\end{figure}

\begin{figure}
\centering
\includegraphics[width=\columnwidth]{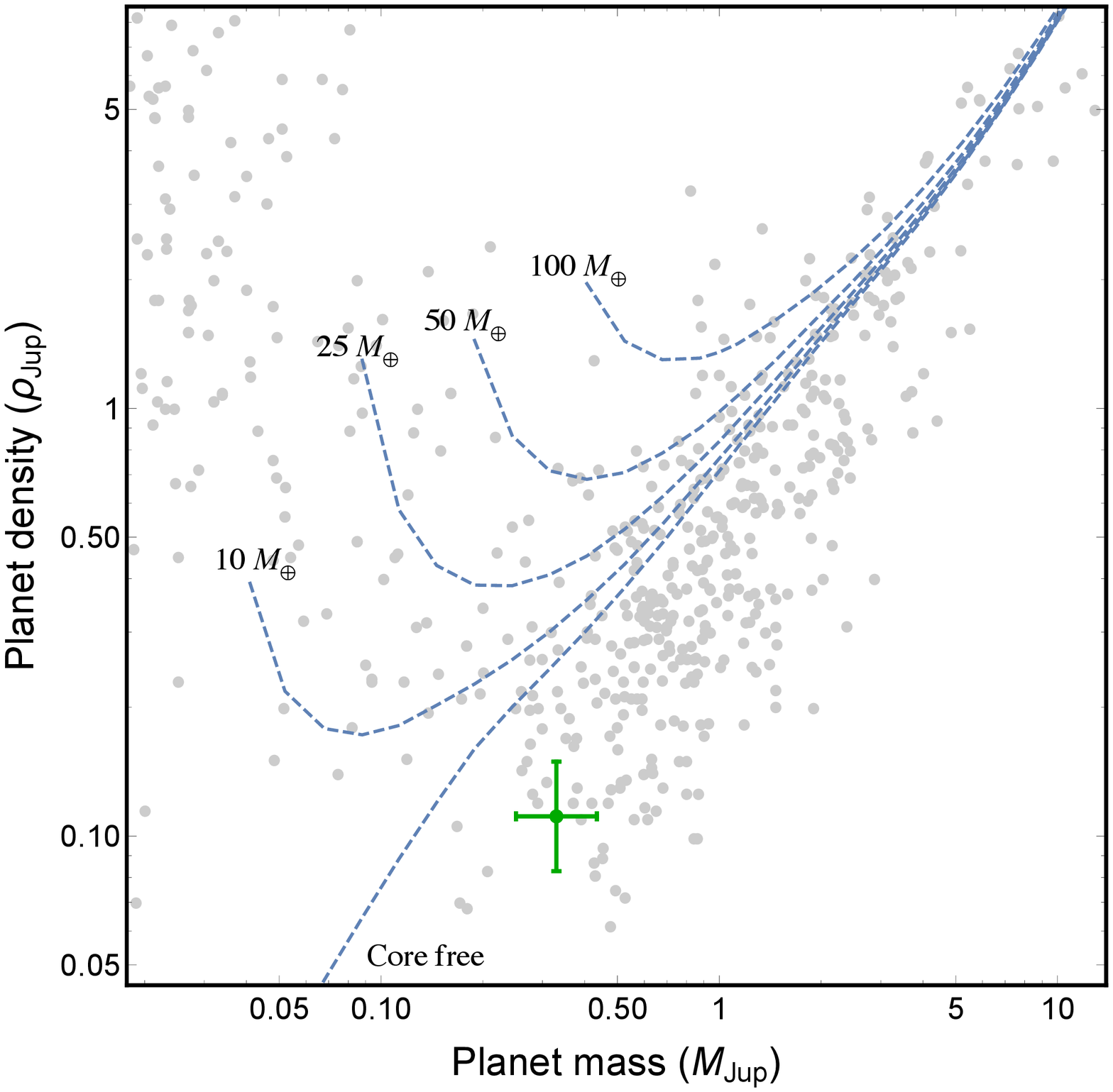}
\caption{
The mass-density diagram of the currently known transiting exoplanets with precisely measured mass and radius. The grey points denote values taken from TEPCat. Their error bars have been suppressed for clarity. The position of WASP-174b is shown in green with error bars. Four planetary models, with various heavy-element core masses (10, 25, 50, and 100 Earth mass) and another without a core are plotted for comparison. They were estimated for a planet at 0.045\,au from a parent star with an age of 3.16\,Gyr \citep{fortney:2007}.
}
\label{fig:mass-density}
\end{figure}

Having measured the RV phase curve of the parent star (see Figure~\ref{fig:RV}), we have also determined that the mass of WASP-174b is $M_{\rm p}=0.330 \pm 0.091 \, M_{\rm Jup}$, which, together with a more precise measurement of its radius, $R_{\rm p}=1.437 \pm 0.050 \, R_{\rm Jup}$, indicates that WASP-174b is an extremely low-density giant planet at the border between Saturn- and Jupiter-type planets (Figure~\ref{fig:mass-period}). The inflated size of WASP-174b is evident from Figure~\ref{fig:mass-radius}. Figures~\ref{fig:radius-gravity} and \ref{fig:mass-density} show its position in the radius-gravity and mass-density diagrams of the currently known transiting exoplanets. In the latter diagram, the position of the planet is compared with five different theoretical models, taken from \citet{fortney:2007}, having different core of heavy-elements: 0, 10, 25, 50, and 100 Earth mass. The models were estimated for a planet at 0.045\,au from a parent star, with an age of 3.16\,Gyr. WASP-174b has a density comparable with the model related to core-free planets.

Since WASP-174b is a very inflated planet, it could be a favorable target for probing its atmosphere through the transmission spectroscopy method. An estimation of the expected signal of its transmission spectrum, $A_{\rm abs}$, can be generally obtained by deriving the contrast in area between the annular region of the planetary atmosphere and that of the star, during a transit event. This can be obtained from the characteristic length scale of the planetary atmosphere \citep[see e.g.][]{bento:2014}, i.e.
\begin{equation}
A_{\rm abs}=\frac{\pi(R_{\rm p}+H)^2-\pi R_{\rm p}^2}{\pi R_{\star}^2}.
\label{absorption_signal}
\end{equation}
In the latter equation $H$ is the pressure scale height of the planetary atmosphere, that is $H = k_{\rm B}T_{\rm eq}/\mu_{\rm m}\,g_{\rm p}$, where $k_{\rm B}$ is Boltzmann's constant and $\mu_{\mathrm{m}}$ is the mean molecular weight, for which the value of $2.3$\,amu is usually adopted for giant-planet atmospheres dominated by H$_2$ and He \citep{lecavelier:2008}. Because of the transits of WASP-174b are grazing, we have to slightly modify Eq.~(\ref{absorption_signal}) for this planet in order to take into account that only a portion of its atmosfere can absorb the light coming from the parent star. This portion is just equal to $A_{\rm ecl}/A_{\rm p}$ and, therefore, we have
\begin{equation}
A_{\rm abs}=\frac{\pi(R_{\rm p}+H)^2-\pi R_{\rm p}^2}{\pi R_{\star}^2} \left(\frac{A_{\rm ecl}}{A_{\rm p}}\right).
\label{absorption_signal_2}
\end{equation}
Assuming that $H \ll R_{\rm p}$, then Eq.~(\ref{absorption_signal_2}) is simply
\begin{equation}
A_{\rm abs}=\frac{2 R_{\rm p}k_{\rm B}T_{\rm eq}}{\mu_{\rm m}\,g_{\rm p}R_{\star}^2}\left(\frac{A_{\rm ecl}}{A_{\rm p}}\right).
\label{absorption_signal_3}
\end{equation}

\begin{figure}
\centering
\includegraphics[width=\columnwidth]{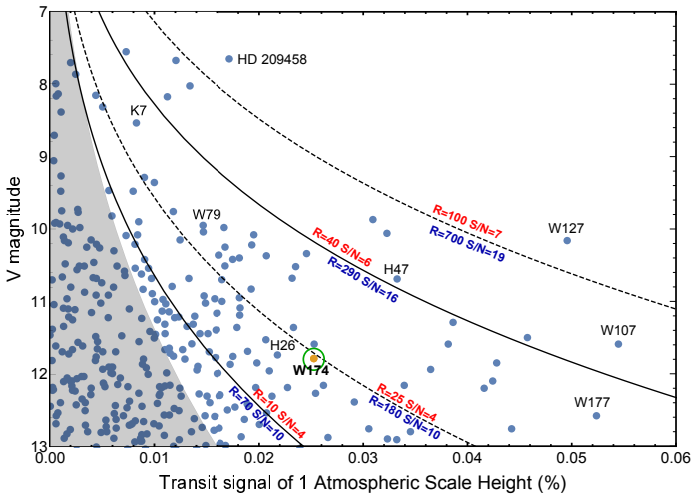}
\caption{Parent-star $V$ magnitude as a function of the expected transmission spectral signal of 1 atmospheric scale height for transiting exoplanets. Data are taken from TEPCat. The curves are adopted from \citet{sing:2018} and give an indication of constant values for S/N and spectral resolution \citep[see also][]{heng:2015}. The numbers in red and blue refer to HST and JWST, respectively. The region of the diagram in which exoplanetary atmospheres are unobservable, due to low S/N, is marked in gray. The position of WASP-174 is highlighted together with those of other selected transiting exoplanets.}
\label{fig:absorption}
\end{figure}

In order to compare the relative signal-to-noise of WASP-174b with those of the other transiting exoplanets, we plotted the absorption signal, $A_{\rm abs}$, against the magnitude of the host stars in Figure~\ref{fig:absorption}. We can see that the expected atmospheric signal for WASP-174b during its transits, in terms of expected S/N, is similar to that of HAT-P-26b, for which evidences of water has been convincingly found with the HST \citep{wakeford:2017}. WASP-174b is, therefore, another interesting close-in giant planet, whose atmospheric composition can be probed via the transmission-spectroscopy technique.

Exploiting our multi-band photometric observations, we attempted to estimate an observational transmission spectrum of WASP-174b. Following the general approach, we made a new fit of each of our complete light curves, which were obtained with the LCGOT/sinistro and the MPG\,2.2~m/GROND facilities, including the NIR data, and fixing the parameters to the values reported in Table~\ref{tab:planetaryparameters}, with the exception of $R_{\rm p}/R_{\star}$, whose uncertainties were estimated by performing 20\,000 Monte Carlo simulations. 

\begin{figure}
\centering
\includegraphics[width=\columnwidth]{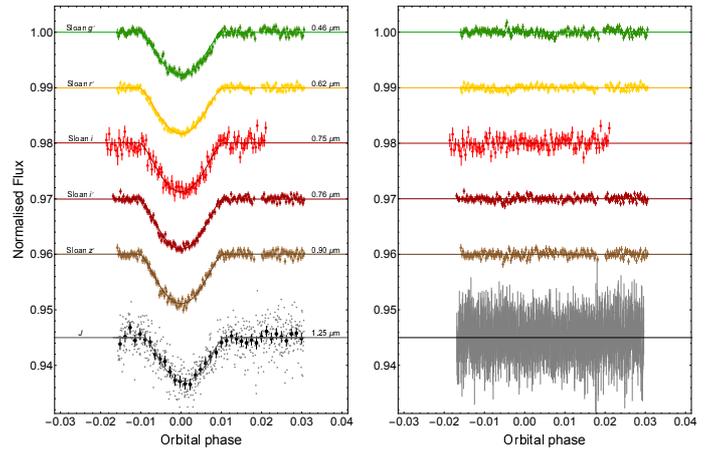}
\caption{{\it Left-hand panel:} the complete light curves recorded by the LCOGT/sinistro ($i$) and MPG\,2.2~m/GROND ($i^{\prime},\,r^{\prime},\,i^{\prime},\,z^{\prime},\,J$) are plotted together with the best-fitting models, which were obtained fixing all the parameters except $R_{\rm p}/R_{\star}$. {\it Right-hand panel:} residual from the best fit models for the corresponding filters.}
\label{fig:lcs_transmission}
\end{figure}

The light curves and the best-fitting models are plotted in Figure~\ref{fig:lcs_transmission}, with the exception of the $H$ and $K$ light curves, because the lower values that we obtained from them are not physically justifiable (the data are simply too noisy to extract reliable results). 

\begin{figure}
\centering
\includegraphics[width=\columnwidth]{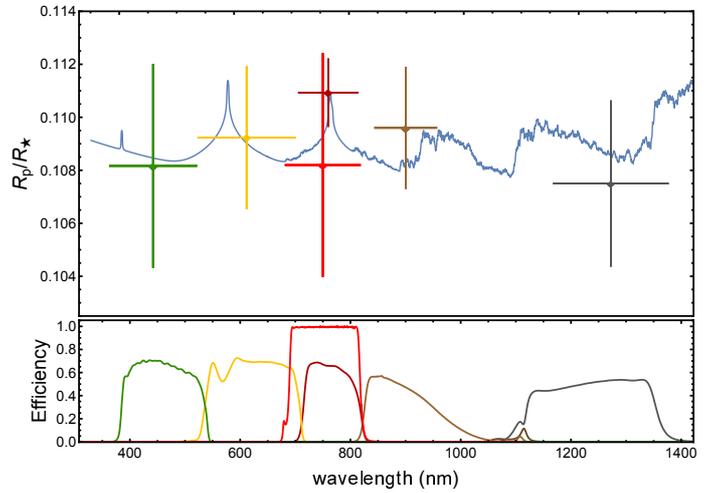}
\caption{Variation of the planet/star radius ratio with wavelength. The colors are the same as in Figure~\ref{fig:lcs}. The vertical bars represent the errors in the measurements and the horizontal bars show the FWHM transmission of the passbands used. The observational points are compared with an atmosphere model from \citet{fortney:2010}. The transmission curve for the Sloan\,$i$ filter and the total efficiencies of the GROND filters are shown in the bottom panel.}
\label{fig:transmission_spectrum}
\end{figure}

The values of $R_{\rm p}/R_{\star}$ are reported in Table~\ref{tab:transmission} and shown in Figure~\ref{fig:transmission_spectrum}, where the vertical errorbars represent the relative errors in the measurements and the horizontal errorbars are related to the FWHM transmission of the passbands used. Just for illustration, we also plotted a model atmosphere calculated by \citet{fortney:2010} for a giant planet with a surface gravity of $g_{\rm p} = 10$\,ms$^{-2}$, a base radius of 1.25 $R_{\rm Jup}$ at 10 bar, and $T_{\rm eq} = 1000$\,K. Clearly, the quality of the data do not allow to characterize the transmission spectrum of WASP-174b and we can only conclude that our observational points do not indicate any large variation of the radius of the planet and are compatible with a flat transmission spectrum. More accurate data are required for effectively studying the atmospheric composition of WASP-174b.

% Table
\begin{table}
\centering
\setlength{\tabcolsep}{4pt}
\caption{Values of $R_{\rm p}/R_{\star}$ for each of the passbands.}
\label{tab:transmission}
\begin{tabular}{lcccccccccccc} 
\hline
Telescope & Filter & $\lambda$ (nm)  & $R_{\rm p}/R_{\star}$  \\
\hline \\[-8pt]
MPG 2.2\,m & Sloan $g^{\prime}$ &  459 &  $0.1082 \pm 0.0038$ \\
MPG 2.2\,m & Sloan $r^{\prime}$ &  622 &  $0.1092 \pm 0.0027$ \\
LCGOT 1\,m & Sloan $i$          &  754 &  $0.1082 \pm 0.0042$ \\
MPG 2.2\,m & Sloan $i^{\prime}$ &  764 &  $0.1109 \pm 0.0013$ \\
MPG 2.2\,m & Sloan $z^{\prime}$ &  899 &  $0.1096 \pm 0.0023$ \\[2 pt]
MPG 2.2\,m & $J$                & 1240 &  $0.1075 \pm 0.0031$ \\
MPG 2.2\,m & $H$                & 1647 &  $0.0996 \pm 0.0029$ \\
MPG 2.2\,m & $K$                & 2170 &  $0.0958 \pm 0.0089$ \\[2 pt]
\hline \end{tabular} %
\end{table}

% Acknowledgements
%-------------------------------------------------------------------
\begin{acknowledgements}
Development of the HATSouth project was funded by NSF MRI grant NSF/AST-0723074, operations have been supported by NASA grants NNX09AB29G, NNX12AH91H, and NNX17AB61G, and follow-up observations have received partial support from grant NSF/AST-1108686. GROND was built by the high-energy group of MPE in collaboration with the LSW Tautenburg and ESO, and is operated as a PI-instrument at the MPG 2.2\,m telescope. The reduced light curves presented in this work will be made available at the CDS (http://cdsweb.u-strasbg.fr/). L.M.\ acknowledges support from the University of Rome ``Tor Vergata'' through ``Mission: Sustainability 2017'' fund.
A.J.\ acknowledges support from FONDECYT project 1171208, and by the Ministry for the Economy, Development, and Tourism's Programa Iniciativa Cient\'{i}fica Milenio through grant IC\,120009, awarded to the Millennium Institute of Astrophysics (MAS). K.P. acknowledges support from NASA grants 80NSSC18K1009 and NNX17AB94G. We thank Jorge Lillo-Box, Elyar Sedaghati and the anonymous referee for their suggestions and comments.
\end{acknowledgements}

%
%-------------------------------------------------------------------


\begin{thebibliography}{}

\bibitem[Bakos et al.(2004)]{bakos:2004} % HATNet survey
Bakos, G.~\'{A}., Noyes, R.~W., Kov\'{a}cs, G., et al. 2004, \pasp, 116, 266 %

\bibitem[Bakos et al.(2013)]{bakos:2013:hatsouth} % HATSouth survey
Bakos G.~\'{A}., Csubry, Z., Penev, K., et al. 2013, \pasp, 125, 154 %

\bibitem[Barentsen et al.(2019)]{barentsen:2019} %
Barentsen, G., Hedges, C., Vin\'{i}cius, Z., et al. 2019, KeplerGO/lightkurve: Lightkurve v1.0b29 (Version v1.0b29), Zenodo

\bibitem[Bento et al.(2014)]{bento:2014} %
Bento, J., Wheatley, P.~J., Copperwheat, C.~M., et al. 2014, \mnras, 437, 1511

\bibitem[Borucki et al.(2010)]{borucki:2010} %
Borucki, W.~J., Koch, D., Basri, G. et al. 2010, Science, 327, 977

\bibitem[Bou{\'e} et~al.(2013)]{boue:2013} %
Bou{\'e}, G., Montalto, M., Boisse, I., et al. 2013, \aap, 550, A53

\bibitem[Bovy et~al.(2016)]{bovy:2016} %
Bovy, J., Rix, H.-W., Green, G.~M., et al. 2016, \apj, 818, 130

\bibitem[Brahm et~al.(2017)]{brahm:2017b} %
Brahm, R., Jord\'{a}n, A., Hartman, J., Bakos, G. 2017, \mnras, 467, 971

\bibitem[Claret(2018)]{claret:2018} %
Claret, A. 2018, \aap, 618, A20

\bibitem[Claret et~al.(2012)]{claret:2012} %
Claret, A., Hauschildt, P.~H., \& Witte, S. 2012, \aap, 546, A14

\bibitem[Claret et~al.(2013)]{claret:2013}
Claret, A., Hauschildt, P.~H., \& Witte, S. 2013, \aap, 552, A16

\bibitem[Cutri et al.(2014)]{cutri:2014} %
Cutri, R.~M., et al.\ 2014, VizieR Online Data Catalog, II/328

\bibitem[Fortney et al.(2007)]{fortney:2007} % 
Fortney J.~J., Marley M.~S., Barnes J.~W., 2007, \apj, 659, 1661

\bibitem[Fortney et al.(2010)]{fortney:2010} % 
Fortney J.~J., Shabram, M., Showman, A.~P., et al. 2010, \apj, 709, 1396

\bibitem[Fressin et al.(2013)]{fressin:2013} % 
Fressin, F., Torres, G., Charbonneau, D., et al. 2013, \apj, 766, 81

\bibitem[Fulton \& Petigura(2018)]{fulton:2018} %
Fulton, B.~J., Petigura, E.~A. 2018, \aj, 156, 264

\bibitem[Greiner et al.(2008)]{greiner:2008} % GROND
Greiner, J., Bornemann, W., Clemens, C., et al. 2008, \pasp, 120, 405

\bibitem[Hartman et al.(2019)]{hartman:2019}
Hartman, J.~D., Bakos, G.~{\'A}., Bayliss, D., et al. 2019, \aj, 157, 55

\bibitem[Hatzes \& Rauer(2015)]{hatzes:2015} %
Hatzes, A.~P., \&  Rauer, H. 2015, \apj, 810, L25

\bibitem[Hellier et al.(2012)]{hellier:2012} % WASP-67
Hellier, C., Anderson, D.~R., Collier Cameron, A., et al. 2012, MNRAS, 426,
739

\bibitem[Heng \& Winn(2015)]{heng:2015}
Heng, K., Winn, J. 2015, American Scientist, 103, 196 

\bibitem[Jenkins et al.(2016)]{jenkins:2016}
Jenkins, J.~M., Twicken, J.~D., McCauliff, S., et al. 2016, \procspie, 9913, 99133E

\bibitem[Knutson et al.(2007)]{knutson:2007} % 
Knutson, H.~A., Charbonneau, D., Noyes, R.~W. 2007, \apj, 655, 564

\bibitem[Lecavelier des Etangs et al.(2008)]{lecavelier:2008} %
Lecavelier des Etangs A., Pont F., Vidal-Madjar A., Sing D., 2008, \aap, 481, L83

\bibitem[Lillo-Box et al.(2015)]{lillo-box:2015} %
Lillo-Box, J., Barrado, D., Santos, N.~C., et al. 2015, \aap, 577, A105

\bibitem[Ma{\'\i}z Apell{\'a}niz \& Weiler(2018)]{maizapellaniz:2018}% 
Ma{\'\i}z Apell{\'a}niz, J., \& Weiler, M.\ 2018, \aap, 619, A180

\bibitem[Mancini et al.(2014)]{mancini:2014} % 
Mancini, L., Southworth, J., Ciceri, S., et al. 2014, \aap, 568, A127

\bibitem[Mancini et al.(2019)]{mancini:2019} % 
Mancini, L., Southworth, J., Molli\'{e}re, P., et al. 2019, \mnras, 485, 5168

\bibitem[Mohler-Fischer et al.(2013)]{mohler:2013} % 
Mohler-Fischer, M., Mancini, L., Hartman, J.~D., et al. 2013, \aap, 558, A55

\bibitem[Marigo et al.(2017)]{marigo:2017} %
Marigo, P., Girardi, L., Bressan, A., et al.\ 2017, \apj, 835, 77

\bibitem[Mayor et al.(2011)]{mayor:2011} %
Mayor, M., Marmier, M., Lovis, C., et al. 2011, arXiv:1109.2497

\bibitem[Pepper et al.(2007)]{pepper:2007} % KELT survey
Pepper, J., Pogge, R.~W., DePoy, D.~L., et al. 2007, \pasp, 119,
923

\bibitem[Perryman et al.(2018)]{perryman:2018} %
Perryman, M. 2018, The Exoplanet Handbook - 2nd edition (Cambridge Univ. Press, Cambridge)

\bibitem[Pollacco et al.(2006)]{pollacco:2006} % SuperWASP
Pollacco, D.~L., Skillen, I., Collier Cameron, A., et al. 2006, \pasp, 118, 1407 %

\bibitem[Ricker et al.(2015)]{ricker:2015} %
Ricker, G.~R., Winn, J.~N., Vanderspek, R., et al. 2015, Journal of Astronomical
Telescopes, Instruments, and Systems, 1, 014003,

\bibitem[Sing(2018)]{sing:2018} %
Sing D., 2018, in Bozza V., Mancini L., Sozzetti A., eds., Astrophysics of Exoplanetary Atmosphere, Springer International Publishing Sing, D. 2018, in Bozza V., Mancini L., Sozzetti A., eds., Astrophysics of Exoplanetary Atmospheres, Astrophysics and Space Science Library. Springer International Publishing, Switzerland, p. 3

\bibitem[Southworth(2011)]{southworth:2011} %
Southworth, J. 2011, \mnras, 417, 2166

\bibitem[Stassun et al.(2018)]{stassun:2018} %
Stassun, K.~G., Oelkers, R.~J., Pepper, J., et al. 2018, \aj, 156, 102 

\bibitem[Temple et al.(2018)]{temple:2018} % WASP-174 discovery paper
Temple, L.~Y., Hellier, C., Almleaky, Y. et al. 2018, \mnras, 480, 5307

\bibitem[Wakeford et al.(2017)]{wakeford:2017}
Wakeford, H.~R., Sing, D.~K., Kataria, T., et al. 2017, Science, 356, 628

\bibitem[Wright et al.(2012)]{wright:2012}
Wright, J.~T., Marcy, G.~W., Howard, A.~W., et al. 2012, \apj, 753, 160

\end{thebibliography}
\end{document}